\documentclass[conference,letterpaper]{IEEEtran}
\IEEEoverridecommandlockouts

\usepackage{graphicx}
\usepackage{xcolor}
\usepackage{booktabs}
\usepackage{caption}
\usepackage{float}
\usepackage{pifont}
\usepackage{cite}
\usepackage{microtype}
\providecommand{\Description}[1]{}%
\newif\ifrevisionhighlights
\revisionhighlightsfalse

\usepackage{newunicodechar}
\newunicodechar{▁}{\textunderscore}
\newunicodechar{ρ}{$\rho$}
\DeclareUnicodeCharacter{00D7}{$\times$}
\DeclareUnicodeCharacter{2014}{---}
\DeclareUnicodeCharacter{2019}{'}
\DeclareUnicodeCharacter{201C}{``}
\DeclareUnicodeCharacter{201D}{''}

\usepackage{amsmath,amssymb,amsfonts}
\usepackage{hyperref}
\usepackage{xurl}%
\usepackage[export]{adjustbox}
\makeatletter
\newcommand{\tablesmaller}{\fontsize{\strip@pt\dimexpr0.9\dimexpr\f@size pt\relax\relax}{\strip@pt\dimexpr1.08\dimexpr\f@size pt\relax\relax}\selectfont}
\makeatother
\setkeys{Gin}{width=\columnwidth, keepaspectratio}
\usepackage{enumitem}
\usepackage{colortbl}
\usepackage[most]{tcolorbox}
\usepackage{pgfplots}

\definecolor{APPONLY}{HTML}{E2F0D9}

\pgfplotsset{compat=1.18}
\usepgfplotslibrary{groupplots,statistics}

\usepackage{pgfplotstable}
\usepackage{algorithm}
\usepackage{algpseudocode}
\usepackage{array}
\usepackage{multirow}
\usepackage{subcaption}
\usepackage{wrapfig}
\usepackage{needspace}
\usepackage{makecell}
\usepackage{tabularx}
\usepackage{seqsplit}
\usepackage{tikz}
\usetikzlibrary{arrows.meta, positioning, matrix, patterns}
\usepackage{listings}
\hypersetup{hidelinks}

\definecolor{bltxt}{HTML}{2A78D6}\definecolor{blimg}{HTML}{EB6834}

\definecolor{DeletionGray}{RGB}{238,238,238}

\definecolor{ClarityGreen}{RGB}{232,247,234}

\definecolor{sig}{HTML}{184F95}\definecolor{ns}{HTML}{A3A29B}\definecolor{ink}{HTML}{0B0B0B}\definecolor{ink2}{HTML}{52514E}\definecolor{grid}{HTML}{E6E5E0}\definecolor{band}{HTML}{F4F3EF}\definecolor{h}{HTML}{6FA8E8}
\definecolor{hotpink}{RGB}{255, 20, 147}

\usepackage{etoolbox}
\AtBeginEnvironment{figure}{\selectcolormodel{gray}}\AtBeginEnvironment{figure*}{\selectcolormodel{gray}}\AtBeginEnvironment{table}{\selectcolormodel{gray}}\AtBeginEnvironment{table*}{\selectcolormodel{gray}}
\pgfplotsset{every axis/.append style={axis line style={black,line width=0.5pt},tick align=inside,xtick pos=both,ytick pos=both,tick style={black,line width=0.4pt},ymajorgrids=false,%
major grid style={densely dashed,black!45,line width=0.3pt},legend style={draw=black,fill=white,line width=0.4pt,rounded corners=0pt},legend cell align=left,tick label style={font=\sffamily\fontsize{6}{7}\selectfont},label style={font=\sffamily\fontsize{6.5}{7.5}\selectfont}}}

\newcommand{\circnum}[1]{(\romannumeral#1\relax)}

\usepackage{placeins}
\IfFileExists{soul.sty}{\usepackage{soul}}{}
\makeatletter
\providecommand{\sethlcolor}[1]{\def\hl@color{#1}}
\providecommand{\hl}[1]{\colorbox{\hl@color}{#1}}
\makeatother
\ifrevisionhighlights
  \colorlet{revhl}{yellow}
\else
  \colorlet{revhl}{white}
  \renewcommand{\hl}[1]{#1}
\fi
\sethlcolor{revhl}

\hypersetup{hidelinks,
  pdftitle={Lasso: Towards Sustainable Real-Time AI-Generated Media Detection with Intel On-Chip Compression Accelerator for the Edge},
  pdfauthor={Darsh Asher, Joshua Kalyanapu, Kaushal Mhapsekar, Bita Aslrousta, Amna Hassan, Debayan Sengupta, Samira Mirbagher Ajorpaz}}

\begin{document}

\title{Lasso: Towards Sustainable Real-Time AI-Generated Media Detection with Intel On-Chip Compression Accelerator \texorpdfstring{for the Edge}{for the Edge}}

\author{\IEEEauthorblockN{Darsh Asher, Joshua Kalyanapu, Kaushal Mhapsekar, Bita Aslrousta,\\
Amna Hassan, Debayan Sengupta, Samira Mirbagher Ajorpaz}
\IEEEauthorblockA{North Carolina State University\\
\{dkasher, jkalyan, kmhapse, baslrou, ahassan9, dsengup2, smirbag\}@ncsu.edu}}

\definecolor{palA}{HTML}{08306B}
\definecolor{palB}{HTML}{2171B5}
\definecolor{palC}{HTML}{6BAED6}
\definecolor{palD}{HTML}{BDD7E7}
\definecolor{palE}{HTML}{EFF3FF}

\renewcommand{\arraystretch}{1.1}   %
\setlength{\heavyrulewidth}{0.08em}
\setlength{\lightrulewidth}{0.04em}
\setlength{\cmidrulewidth}{0.03em}

\newcommand{\scheme}{Lasso}
\newcommand{\qatscan}{Lasso}
\newcommand{\lassoX}{Lasso-Extract}
\newcommand{\lassoF}{Lasso-Fingerprint}
\newcommand{\lassoD}{Lasso-Detect}
\newcommand{\lasso}{Lasso}

\definecolor{LNAVY}{HTML}{1F3E5C}
\definecolor{LRED}{HTML}{A33240}
\definecolor{LGRAY}{HTML}{8A8A8A}
\definecolor{LGRAYL}{HTML}{C8C8C8}

\pgfplotsset{
  every axis/.append style={
     axis line style={LNAVY!85, line width=0.5pt},
     tick style={LNAVY!85, line width=0.5pt},
     label style={font=\small},
     tick label style={font=\footnotesize},
     grid=none, ymajorgrids=false, xmajorgrids=false,
  },
  every axis legend/.append style={draw=none, fill=none, font=\footnotesize},
  cycle list={{LNAVY}, {LRED}, {LGRAY}, {LNAVY, dashed}, {LGRAY, dashed}},
}

\providecommand{\dsetup}{%
  \scriptsize\setlength{\tabcolsep}{3pt}\renewcommand{\arraystretch}{1.12}\arrayrulecolor{LNAVY!45}}
\providecommand{\cnt}[4]{\par{\tiny\color{LNAVY!85!black}train #1, #2; test #3, #4}}
\providecommand{\Cx}[1]{#1}
\providecommand{\Sx}[1]{\par#1}
\providecommand{\frm}[4]{\par{\tiny\bfseries\color{orange!75!black}frames: train #1, #2; test #3, #4}}
\providecommand{\thpair}[2]{\includegraphics[height=0.75cm,valign=t]{dataset_examples/#1}\,\includegraphics[height=0.75cm,valign=t]{dataset_examples/#2}}
\providecommand{\dhead}{\textbf{Dataset} {\tiny(train H, AI; test H, AI)} & \textbf{Complexity and significance} & \textbf{Human vs.\ AI example} \\ \toprule}

\newcommand{\cmark}{\textcolor{LNAVY}{\ding{51}}}
\newcommand{\lcheck}{\textcolor{LRED}{\ding{51}}}
\newcommand{\xmark}{\textcolor{LGRAYL}{\ding{55}}}
\newcommand{\nmark}{\textcolor{LGRAY}{}}
\DeclareRobustCommand{\incnote}[1]{\textcolor{blue}{\footnotesize\,[#1]}}

\lstdefinestyle{codestyle}{
  basicstyle=\ttfamily\footnotesize, breaklines=true, showstringspaces=false,
  columns=fullflexible, keepspaces=true, language=Python, frame=none, upquote=true
}

\maketitle
\thispagestyle{plain}
\pagestyle{plain}

\begin{abstract}
Synthetic text, images, and video are rising. 
Software-only detectors are power-hungry, too slow, and costly, and do not scale with demand. When software cannot scale, historically hardware often comes to the rescue.
We perform the first reverse engineering of a compression accelerator, the
Intel QuickAssist Technology (QAT), and find that its latency depends on how
its input repeats, not on what the bytes are, a property we call
\textit{repetition-structure dependence}.
Using this information, we then build \scheme{}, a low-cost, high-performance detector for
 AI-generated content that turns the accelerator into a measuring
 instrument.

For text detection, \scheme{} achieves $100\%$ accuracy on original DetectRL text and retains $92\%$ average accuracy after paraphrasing, compared with $78\%$ for Binoculars and $55\%$ for DetectGPT. Across original, paraphrased, and lightly perturbed text, where
perturbation inserts typos and substitutes words, it achieves $82\%$
average accuracy, outperforming all five baselines. With topic, length, and formatting matched, accuracy reaches $95\%$ for instruction-tuned generators, including when the tested generator is unseen. On the RAID benchmark of $11$ text generators and eight writing domains, accuracy averages $82\%$ when trained on one generator and tested on another, and $74\%$ across eight held-out domains. Detection takes $197$\,$\mu$s, at least $356\times$ faster than each of the five baselines. At the same time, \scheme{} consumes over $99\%$ less energy per check than both Binoculars and DetectGPT.

Beyond text, \scheme{} detects AI-generated videos, software code, and images with 78\%, 70\% and 98\% accuracy, respectively, with low latency and power opening avenues for practical real time and edge device detection of multimodality AI artifacts for the first time. \hl{No other compared technique supports detection across all four modalities: text, video, images, and code.}
\end{abstract}
\section{Introduction}
\label{sec:intro}

Huawei projects that the world's nine billion people will be connected to 900 billion AI agents by 2035~\cite{huawei2035}. 
Each of these agents produces text, images, code, and video as it works, so
the volume of AI-generated artifacts will keep growing, and every new artifact
adds another item whose origin must be verified. Graphite, a web-analytics company that samples published web pages, estimates about half of newly published online articles are AI-generated~\cite{ref:graphite}. The line between synthetic and authentic media has grown difficult for people to draw reliably~\cite{jiang2026wrath}.
The consequences of this gap are no longer hypothetical:
AI-generated content has already been implicated in election interference, financial fraud, and litigation~\cite{nhvotersuppression2024, arup_deepfake2024, yang_v_umn2025, doe_v_yale2025, harris_v_hingham2024, zelenskyy_deepfake2022}. The question is no longer whether \emph{provenance detection}, that is, whether a piece of
data is human- or machine-produced, matters, but whether it can \emph{scale}: fast, cheap, and deployable at the point where content is consumed, keeping pace with content that grows with the generative model's size and accuracy.

 \begin{figure}[t]
\centering
\selectcolormodel{natural}\definecolor{lassoRed}{HTML}{C44E52}%
\resizebox{\columnwidth}{!}{%
\begin{tikzpicture}[x=1pt,y=1pt,font=\fontsize{6.5}{7.2}\selectfont]
\node[anchor=east,font=\bfseries\fontsize{6.5}{7.2}\selectfont] at (53,96) {\scheme{}};
\node[anchor=east] at (53,83) {Log-Rank};
\node[anchor=east] at (53,70) {Fast-DetectGPT};
\node[anchor=east] at (53,57) {Binoculars};
\node[anchor=east] at (53,44) {DetectGPT};
\node[anchor=east] at (53,31) {DNA-GPT};
\node[font=\bfseries\fontsize{6.5}{7.2}\selectfont] at (86,120) {(a) Energy};
\draw[black,line width=.35pt] (58,20)--(114,20);
\draw[line width=.35pt] (58,20)--(58,18);\node[anchor=north,font=\fontsize{5.8}{6.5}\selectfont] at (58,17) {$10^{-5}$};
\draw[line width=.35pt] (76.66666666666667,20)--(76.66666666666667,18);\node[anchor=north,font=\fontsize{5.8}{6.5}\selectfont] at (76.66666666666667,17) {0.01};
\draw[line width=.35pt] (95.33333333333334,20)--(95.33333333333334,18);\node[anchor=north,font=\fontsize{5.8}{6.5}\selectfont] at (95.33333333333334,17) {10};
\draw[line width=.35pt] (114,20)--(114,18);\node[anchor=north,font=\fontsize{5.8}{6.5}\selectfont] at (114,17) {10k};
\node at (86,1) {J/check};
\fill[lassoRed] (58,93) rectangle (71.02,99);
\fill[black!58.5] (58,80) rectangle (96.65,86);
\fill[black!58.5] (58,67) rectangle (96.65,73);
\fill[black!58.5] (58,54) rectangle (97.80,60);
\fill[black!58.5] (58,41) rectangle (110.97,47);
\fill[black!58.5] (58,28) rectangle (105.08,34);
\draw[black,densely dashed,line width=.65pt] (87.24,20)--(87.24,100);
\node[anchor=south,font=\fontsize{5.8}{6.5}\selectfont,fill=white,inner sep=.4pt] at (86,100) {\shortstack{Edge-device power budget\\5--6 W at 10 checks/s}};
\node[font=\bfseries\fontsize{6.5}{7.2}\selectfont] at (157,120) {(b) Latency};
\draw[black,line width=.35pt] (129,20)--(185,20);
\draw[line width=.35pt] (129,20)--(129,18);\node[anchor=north,font=\fontsize{5.8}{6.5}\selectfont] at (129,17) {0.1};
\draw[line width=.35pt] (147.66666666666666,20)--(147.66666666666666,18);\node[anchor=north,font=\fontsize{5.8}{6.5}\selectfont] at (147.66666666666666,17) {100};
\draw[line width=.35pt] (166.33333333333334,20)--(166.33333333333334,18);\node[anchor=north,font=\fontsize{5.8}{6.5}\selectfont] at (166.33333333333334,17) {100k};
\draw[line width=.35pt] (185,20)--(185,18);\node[anchor=north,font=\fontsize{5.8}{6.5}\selectfont] at (185,17) {100M};
\node at (157,1) {\textmu{}s/check};
\fill[lassoRed] (129,93) rectangle (149.50,99);
\fill[black!58.5] (129,80) rectangle (165.37,86);
\fill[black!58.5] (129,67) rectangle (165.41,73);
\fill[black!58.5] (129,54) rectangle (168.63,60);
\fill[black!58.5] (129,41) rectangle (180.27,47);
\fill[black!58.5] (129,28) rectangle (176.13,34);
\draw[black,densely dashed,line width=.65pt] (166.3333,20)--(166.3333,100);
\node[anchor=south,font=\fontsize{5.8}{6.5}\selectfont,fill=white,inner sep=.4pt] at (166.3333,100) {Real-time: 100 ms};
\node[font=\bfseries\fontsize{6.5}{7.2}\selectfont] at (228,120) {(c) Robustness};
\draw[black,line width=.35pt] (200,20)--(256,20);
\draw[line width=.35pt] (200,20)--(200,18);\node[anchor=north,font=\fontsize{5.8}{6.5}\selectfont] at (200,17) {0};
\draw[line width=.35pt] (228,20)--(228,18);\node[anchor=north,font=\fontsize{5.8}{6.5}\selectfont] at (228,17) {0.5};
\draw[line width=.35pt] (256,20)--(256,18);\node[anchor=north,font=\fontsize{5.8}{6.5}\selectfont] at (256,17) {1};
\node at (228,1) {Paraphrase AUROC};
\fill[lassoRed] (200,93) rectangle (254.376,99);
\fill[black!58.5] (200,80) rectangle (237.912,86);
\fill[black!58.5] (200,67) rectangle (240.992,73);
\fill[black!58.5] (200,54) rectangle (245.248,60);
\fill[black!58.5] (200,41) rectangle (225.872,47);
\fill[black!58.5] (200,28) rectangle (241.552,34);
\end{tikzpicture}%
}
\caption{\scheme{} on DetectRL: lowest energy and latency, top AUROC.}
\label{fig:intro-energy-roc}
\label{fig:performance}
\end{figure}
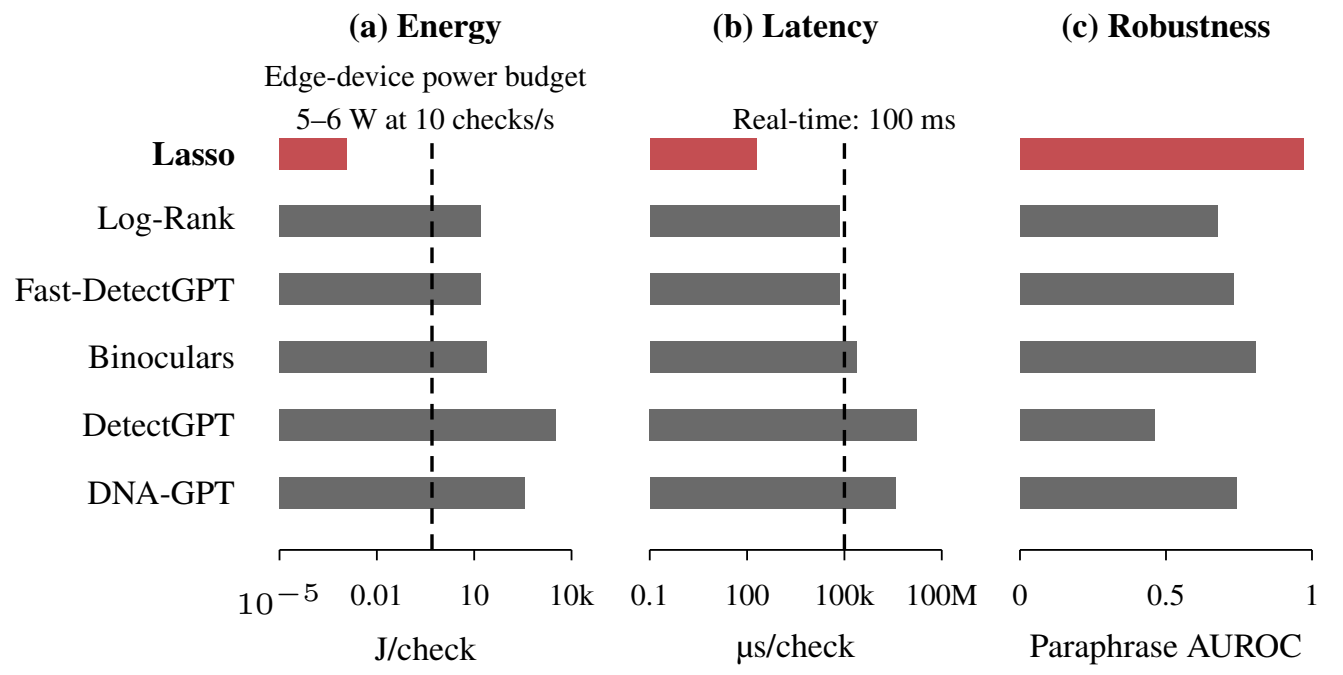

Defenses to date are almost entirely software: every check runs a learned
detector on general-purpose hardware, at a cost that grows with both the
detector's size and the artifact's size. A $355$M-parameter RoBERTa
classifier spends about $8$\,J per article~\cite{ref:solaiman}, Binoculars
runs two $7$B-parameter LLMs per article for $24.9$\,J~\cite{hans2024binoculars},
and one minute of video costs $432$\,J with a frame classifier and
$72{,}000$\,J with a detector that regenerates each frame with a diffusion
model~\cite{ref:dire}. Checking the $2\times10^{8}$ minutes of video uploaded
daily~\cite{ref:youtube} would thus need $2\times10^{4}$ to $5\times10^{5}$
H100-class GPUs running continuously, or \$$0.4$B to \$$11$B per year at
\$$2.3$ per GPU-hour. These detectors are also too large and power-hungry for the phones and edge devices where content is created: Binoculars' two models need about $28$\,GB of 16-bit weights, more
than a phone or edge device holds, and every software detector exceeds a
$5$--$6$\,W edge power budget at ten checks per second
(Figure~\ref{fig:intro-energy-roc}(a)). They fail under simple evasion: on
paraphrased DetectRL text, their AUROC is only $0.46$--$0.81$, with DetectGPT
below chance~\cite{wu2024detectrl} (Figure~\ref{fig:intro-energy-roc}(c)).
Each reads a single modality, so none can check a file that mixes them, such
as an image with embedded text. Finally, their features are visible to
software, so an attacker can observe and shape the signal they
use~\cite{mirbagher2020perspectron, mirbagher2022evax, khasawneh2017rhmd}.
Therefore, software alone cannot keep pace with AI-generated content.

Little attention has been paid to hardware solutions for data
\emph{provenance} that save power and scale with this demand.
The closest hardware solution, Deepfake CLI~\cite{bhilare2022deepfakecli},
runs quantized neural-network detectors on an FPGA card that requires a
225\,W power budget~\cite{amd2023vck5000power}; thus, it remains power-hungry.
Moreover, CLI accelerates an existing detector and does not ask whether
computation leaves a distinct, measurable microarchitectural footprint that
differentiates AI-generated data from real data.
This paper takes the first step by examining such a footprint on Intel QuickAssist
Technology (QAT), a compression accelerator already integrated into server and edge processors~\cite{intel-xeon-d}.

QAT is a built-in accelerator for compression and cryptographic operations~\cite{intel_what_is_qat}. QAT implements DEFLATE, the lossless compression algorithm behind gzip and zip files.
DEFLATE shrinks data by replacing repeats: it scans the input, and whenever a byte sequence has already appeared earlier, it replaces the repeat with a short pointer back to the earlier copy.
Its output is therefore a stream of two kinds of \emph{tokens}: a \emph{literal}, a single byte copied through unchanged because no earlier repeat was found, and a \emph{back-reference}, an instruction of the form ``copy $L$ bytes from $D$ bytes back''; the $L$ bytes a back-reference copies are its \emph{matched bytes}.
For example, in \emph{the cat sat on the mat}, the second \emph{the} can be encoded as a back-reference to the first.

\pgfplotstableread[header=false]{
6.7386
6.9181
6.9362
6.9700
6.9437
6.9430
6.9279
6.8509
6.9805
6.9839
6.9914
6.9097
7.0370
6.9945
6.9454
6.9720
6.9306
6.9867
6.9679
6.9741
6.9789
6.9455
6.9550
6.9644
6.9799
6.9768
6.9660
6.9289
6.9581
6.8848
6.9685
6.9739
6.9965
6.9551
6.9891
6.8544
6.9680
6.9269
6.9819
6.9920
6.9933
6.9834
6.9582
6.9948
6.9955
6.9733
6.9538
6.9895
6.9992
7.0331
6.9576
6.9871
6.9526
7.0025
6.9535
6.9676
6.9811
6.9572
6.9725
6.9538
7.0222
6.9710
6.9629
7.0386
6.9940
6.9655
6.9912
6.9738
6.9504
6.9895
7.0106
6.9856
6.8948
6.9561
6.9789
6.9986
6.9670
6.9997
6.9351
7.0319
6.9397
7.0102
7.0024
6.9888
6.9743
6.9406
7.0450
6.9351
7.0311
6.9819
6.9355
6.9846
6.9422
6.8333
6.9156
6.9355
6.9390
6.9803
6.9787
6.9810
6.9391
6.9187
6.9588
6.9604
6.9939
6.9244
6.9823
6.9498
6.9712
6.9890
6.9156
6.9258
6.9418
7.0379
7.0283
6.9635
6.9867
6.9278
6.9906
6.9227
6.9761
6.9585
7.0039
6.9893
6.9561
6.8949
6.9664
6.9580
6.9366
7.0102
6.9203
6.9966
6.9908
6.9681
6.9757
6.9793
6.9860
6.9492
6.9671
6.9998
6.9537
6.9822
6.9646
6.9977
6.9684
6.9486
6.9592
6.9159
6.9689
6.9529
6.9188
6.9520
6.9289
6.9208
6.9672
6.9714
6.9855
6.9832
6.9880
6.9585
6.9707
6.9594
6.9634
6.9596
6.9961
6.9544
6.9609
6.9291
6.9305
6.9286
6.9236
7.0011
6.9690
6.9340
6.9979
6.9611
6.9408
6.9461
6.9739
6.9584
6.9870
6.9637
6.9651
6.9197
6.9626
6.9753
6.9956
6.9481
6.9467
6.9984
6.9640
6.9841
7.0234
6.9397
6.9620
6.9680
7.0074
6.9834
6.9949
7.0477
6.9799
6.9121
6.8408
6.9177
6.9873
7.0007
6.9923
6.9573
6.8389
6.9508
6.9697
6.9563
6.9899
7.0051
6.9447
6.9992
6.9556
6.9487
7.0008
}\cpuHuman

\pgfplotstableread[header=false]{
6.9651
6.9691
6.9327
6.9686
6.8417
6.9648
6.9621
6.9629
6.9586
6.9543
6.9350
6.9619
6.9675
6.9705
6.9197
6.9175
6.9205
6.9422
6.9690
6.9846
6.9212
6.9126
6.9781
6.9117
6.9227
6.9751
6.9663
6.9431
6.9291
7.0058
7.0043
6.9784
7.0199
6.9545
6.9874
6.9984
6.9652
6.9549
7.0022
6.7203
6.9801
6.9846
7.0137
6.9943
6.9273
7.0099
6.9774
6.9663
6.9740
6.9790
6.9726
6.9417
6.9566
6.9916
7.0133
6.9771
6.9435
7.0067
6.9996
6.9674
6.9867
6.9449
6.9695
6.9579
6.9823
6.9940
7.0328
6.9284
6.9766
6.9373
6.9530
6.9480
7.0218
6.9803
6.9489
7.0391
6.9334
7.0164
6.9261
6.9554
6.9614
6.9409
6.9863
6.9673
6.9334
6.9871
6.9928
6.9621
6.9817
6.9670
6.9056
6.9835
6.9173
6.9957
6.9546
6.9406
6.9892
6.9600
6.9440
6.9502
6.9704
6.9829
6.9289
6.9860
6.9537
6.9200
6.9824
6.9076
6.9871
6.9434
6.9566
6.9702
6.9222
6.9360
6.9649
6.9902
6.9211
6.9326
6.9440
6.9932
6.9844
7.0025
6.9924
6.9543
6.9131
6.9666
6.9704
6.9315
6.9500
6.9345
6.8779
6.9581
6.9217
7.0286
6.9544
6.9666
6.9909
6.9480
6.9532
6.9898
6.9434
6.9884
6.9606
6.9580
6.9936
6.9135
6.9626
6.9412
6.9984
7.0052
6.9864
6.9627
6.9795
6.9766
6.9743
6.9272
6.9984
6.9623
6.9662
6.9395
6.9345
6.9865
6.9977
6.9602
6.9857
7.0418
6.9392
6.9871
7.0029
7.0075
6.9754
6.9793
6.9803
}\cpuAI

\pgfplotstableread[header=false]{
25.6115
25.7832
25.8111
25.5997
27.8025
25.7374
25.6568
25.8284
26.8384
26.6596
25.3402
26.2151
25.3678
27.6197
26.7101
25.3030
27.1754
25.7546
26.4711
25.3222
26.1773
26.3671
25.6069
26.7938
25.3730
25.3421
25.1604
26.3470
26.6546
25.8124
25.1427
27.6289
25.7281
25.5552
26.9360
25.5200
26.7394
25.3511
28.6812
27.5657
25.8585
26.0132
25.3113
28.3174
25.3158
26.2192
26.3366
24.7182
26.5329
26.4788
25.3357
25.5434
28.7249
25.2029
24.9819
25.1395
25.8331
28.1873
25.3627
25.5460
25.1421
26.0546
26.1185
26.2715
25.3241
26.3810
26.4564
26.0031
25.0834
25.7288
25.3826
25.8311
25.4758
27.6442
25.7116
25.3711
25.5265
25.7169
26.6681
26.7938
26.2364
26.4767
25.8318
25.5415
26.3283
25.5480
26.8644
25.8264
25.5323
26.7165
25.5938
25.1244
26.0166
26.3845
27.6159
26.8774
24.9258
25.5938
26.0031
26.4180
25.3248
25.5990
25.9572
25.1181
26.4075
26.6276
26.4285
25.8045
26.6510
25.7672
25.3962
25.3421
26.2302
25.1345
25.8104
25.5310
25.1357
25.0620
25.3518
27.3060
25.8231
25.3897
25.3011
25.5715
26.4222
25.9282
27.0651
26.0376
25.1541
29.8320
25.3878
25.3434
26.4347
26.6880
25.1838
25.3235
25.9605
25.6253
25.8278
25.3107
25.9020
25.3833
26.0261
27.0570
25.6937
25.9686
26.0722
25.3376
25.1832
25.0639
26.3491
25.9828
26.6461
24.9128
25.6226
26.0423
25.3075
26.0268
25.6753
25.3389
27.6144
25.7162
26.1595
25.5925
25.7666
26.7301
27.5953
28.4422
25.1338
27.6817
26.4943
26.6866
24.9047
25.3511
25.3627
25.8511
26.9818
25.8071
25.3280
25.3807
25.5441
25.4926
25.3588
25.8311
25.8371
25.9794
25.7533
26.9571
25.3530
25.9047
25.1124
25.3627
25.3530
26.9331
26.7073
25.7387
28.0773
25.5611
25.1383
26.8075
26.4592
25.9970
26.5506
25.9565
25.1054
25.5558
25.6233
26.0376
26.0166
25.3376
25.7460
26.5569
27.9517
25.5428
28.8567
25.9774
27.3306
25.3306
25.8378
}\qatHuman

\pgfplotstableread[header=false]{
27.3748
27.3853
29.0706
27.7285
28.7439
29.2877
27.9830
27.9994
32.3279
28.7323
27.6717
27.3845
27.7277
28.2326
28.2470
28.2406
28.1817
29.3488
28.1825
27.6970
29.7000
29.2886
28.2662
30.6429
27.6817
28.6295
28.1873
28.9578
27.7870
27.1577
27.3665
27.4085
27.7331
28.8101
28.4852
28.3632
26.8276
27.6993
27.7462
27.7446
27.4401
27.1592
28.9545
29.3488
27.4371
28.2893
31.0424
27.4484
28.1706
28.3246
27.4454
27.7832
27.8071
27.4620
28.3680
28.6066
27.3441
27.5429
29.7442
27.7146
27.5664
27.6955
27.9767
28.1944
28.8625
28.8234
27.7662
27.4213
27.7300
28.4819
28.1690
27.4801
27.1621
27.3793
27.4567
27.8559
29.5168
28.4560
29.2783
28.0403
29.6235
28.1785
27.9486
28.4722
28.1714
27.7446
28.5168
27.7924
29.1282
28.1865
28.0340
30.1595
28.0316
28.8500
28.7828
29.0318
27.9384
31.3293
28.4414
28.5404
28.2526
28.6451
28.6895
28.1746
28.2901
27.9337
27.4748
28.0584
27.4869
28.7621
28.4698
27.9822
27.6671
28.8675
28.8517
27.8621
27.9822
30.2939
29.9034
27.8211
31.6336
27.8017
28.7770
28.5120
26.8810
28.4698
28.4876
27.6625
27.4869
27.4115
27.4567
29.6991
28.7911
28.9863
27.3935
28.4260
27.4642
28.5829
28.4730
28.8184
27.7654
27.1488
27.5255
28.5160
28.1397
28.1334
28.7819
27.0937
27.6518
28.5120
28.4860
27.3411
28.0513
27.7185
28.0568
30.2206
27.4258
28.2079
27.4801
27.9080
27.9650
28.0403
28.6730
27.9767
27.4748
28.0222
27.6870
27.9166
28.6033
28.5356
27.1076
27.4469
27.5035
}\qatAI

\definecolor{clsHuman}{HTML}{8FAFCF}\definecolor{clsHumanEdge}{HTML}{34526E}
\definecolor{clsAI}{HTML}{C9787C}\definecolor{clsAIEdge}{HTML}{7A2E33}
\begin{figure}[!t]
\centering\selectcolormodel{natural}
\begin{tikzpicture}
\begin{groupplot}[
    group style={group size=2 by 1, horizontal sep=1.3cm},
    width=0.46\linewidth, height=2.65cm,
    ylabel={Count},
    tick label style={font=\small}, label style={font=\small},
    title style={font=\small, yshift=-1ex},
    legend style={font=\small, draw=none, fill=none},
    axis line style={black!80,line width=0.5pt}, tick style={black!80}, xtick pos=bottom, ytick pos=left,
    scaled x ticks=false,
]
\nextgroupplot[title={\textbf{CPU}\enspace AUROC 0.51}, xlabel={Monitor throughput (k jobs/s)}, xmin=6.707, xmax=7.061, ymin=0, ymax=45,
    legend columns=2, legend cell align=left,
    legend style={at={(1.25,1.86)}, anchor=north, /tikz/every even column/.append style={column sep=8pt}},
]
\addplot+[hist={bins=30,data min=6.707,data max=7.061}, draw=clsHumanEdge, line width=0.3pt, fill=clsHuman,
    fill opacity=0.75, mark=none, area legend] table[y index=0] {\cpuHuman};
\addlegendentry{Contending with human}
\addplot+[hist={bins=30,data min=6.707,data max=7.061}, draw=clsAIEdge, line width=0.3pt, fill=clsAI,
    fill opacity=0.75, mark=none, area legend] table[y index=0] {\cpuAI};
\addlegendentry{Contending with AI}
\addplot[forget plot, dashed, semithick, clsHumanEdge] coordinates {(6.9676,0) (6.9676,45)};
\addplot[forget plot, dashed, semithick, clsAIEdge] coordinates {(6.9663,0) (6.9663,45)};
\nextgroupplot[title={\textbf{QAT}\enspace AUROC 0.96}, xlabel={Monitor throughput (k jobs/s)}, xmin=24.414, xmax=32.632, ymin=0, ymax=50,
]
\addplot+[hist={bins=30,data min=24.414,data max=32.632}, draw=clsHumanEdge, line width=0.3pt, fill=clsHuman,
    fill opacity=0.75, mark=none, area legend] table[y index=0] {\qatHuman};
\addplot+[hist={bins=30,data min=24.414,data max=32.632}, draw=clsAIEdge, line width=0.3pt, fill=clsAI,
    fill opacity=0.75, mark=none, area legend] table[y index=0] {\qatAI};
\addplot[forget plot, dashed, semithick, clsHumanEdge] coordinates {(25.8231,0) (25.8231,50)};
\addplot[forget plot, dashed, semithick, clsAIEdge] coordinates {(28.0584,0) (28.0584,50)};
\end{groupplot}
\end{tikzpicture}
\caption{Monitor throughput on another core, lower if a bundle holds QAT longer: QAT separates 392 human/AI bundles; CPU zlib does not.
}
\label{fig:detectrl_cpu_qat}
\end{figure}
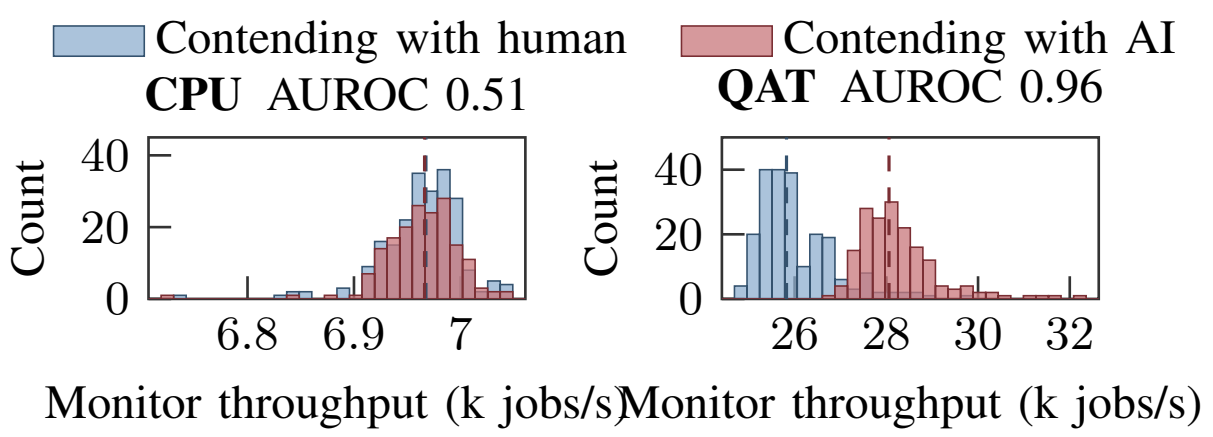

We observe that when we run the same compression algorithm on an isolated general-purpose CPU and on QAT for human text and DIPPER-paraphrased AI text, CPU-based compression barely changes throughput between human and AI text (Figure~\ref{fig:detectrl_cpu_qat}). However, QAT compression creates a distinct microarchitectural footprint between human and AI throughput, showing that the accelerator does not merely reveal a pre-existing software
feature, but it \emph{creates a novel provenance channel}.

\begin{figure}[!t]
  \centering
  \includegraphics[width=0.8\columnwidth]{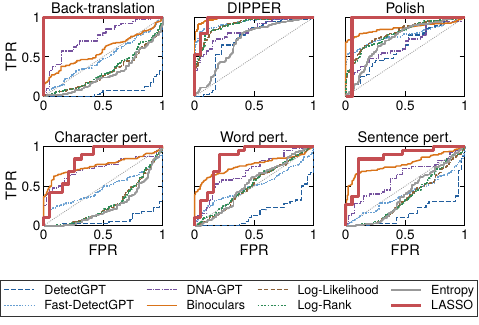}
  \caption{ROC under six rewriting attacks; threshold fixed on unedited text.}
  \label{fig:roc-sweep}
\end{figure}

\begin{figure}[!t]
\centering\begingroup
\pgfplotsset{modalityaxis/.style={
width=\linewidth,height=2.7cm,
axis line style={black,line width=0.5pt},tick style={black,line width=0.4pt},
label style={font=\fontsize{6}{7}\selectfont,color=black},
tick label style={font=\fontsize{6}{7}\selectfont,color=black},
ymajorgrids=false,clip=false,%
}}
\begin{minipage}{0.62\columnwidth}\centering
\begin{tikzpicture}
\begin{axis}[modalityaxis,xmode=log,log basis x=2,
xtick={512,16384,262144},xticklabels={0.5,16,256},
xlabel={Text input (KB)},ylabel={Accuracy (\%)},
ymin=60,ymax=103,ytick={60,80,100},xminorticks=false,enlarge x limits=0.06]
\addplot[black,line width=0.8pt,mark=*,mark size=1.6pt] coordinates {(512,67.2) (4096,78.0) (16384,83.0) (65536,85.9) (131072,93.0) (262144,98.0)};
\end{axis}\end{tikzpicture}
\end{minipage}
\endgroup
\caption{WikiText/GPT-2 detection accuracy vs. text input size.}
\label{fig:observation-budget}\label{fig:prelim-silicon}
\end{figure}
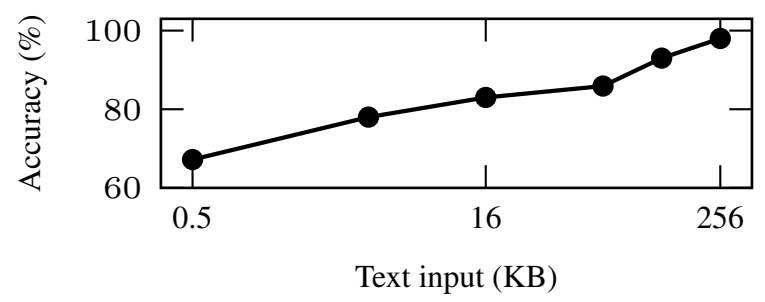

Based on this, we build \scheme{}, a lightweight detector that submits the
artifact under test to QAT as an ordinary compression request while a \emph{monitor process} on another core shares QAT, and passes the monitor's throughput and the artifact's latency to a simple classifier (logistic regression for text and code, gradient-boosted trees for images and video). A shorter artifact compression lets the monitor start sooner, which raises its throughput.
Because paraphrasing changes words but leaves much of the repetition
structure in place, the signal survives evasion: DIPPER~\cite{krishna2023dipper}
rewrites on DetectRL keep the meaning (median embedding
similarity 0.97) while changing much of the wording.
With its threshold fixed on unedited text, \scheme{} still separates human
from AI text under six attacks that either rewrite the text with an LLM or
perturb it with small, meaning-preserving edits.
(Figure~\ref{fig:roc-sweep}), and against a detector-aware adaptive attacker
its resilience is comparable to that of prior detectors
(Figure~\ref{fig:adaptive}).
Larger inputs give QAT more local repeats to measure: on WikiText/GPT-2,
accuracy rises from 67\% at 512\,B to 98\% at 256\,KB
(Figure~\ref{fig:prelim-silicon}).

To investigate why Intel QAT can be used for data provenance, we perform the
first reverse engineering of QAT that asks not how to leak data through it but
how its timing reveals the origin of the data it processes, and we uncover a
previously undocumented dependence of its latency on the repetition structure
of its input.
QAT's latency follows neither the entropy of its input, which measures how
unpredictable its individual bytes are, nor its compression ratio. English text
has more entropy than an all-zero input (4.3 vs.\ 0 bits/byte) and
compresses $7\times$ worse ($12\times$ vs.\ $81\times$), yet completes
$1.5\times$ faster (Table~\ref{tab:char-backrefs}).
Instead, three stages of its match finder set the cost.
A candidate table of limited capacity decides which repeats are found, so
only repeats that are recent and close enough survive; a 48-byte comparator
confirms any shorter match at one fixed cost, whatever its length; and a
separate extender completes longer matches at extra cost.
QAT's latency is therefore a weighted count of the bytes it must emit as
literals and of the matches it finds.

Model-based detectors score text after tokenization~\cite{mitchell2023detectgpt, hans2024binoculars} or images after decoding and resizing~\cite{zhou2026beyondsemantic}, so the byte-level token rate never reaches their input. Compression-based software detectors~\cite{aidetx,lv2026compression} measure output size, which reflects the token count but not the time the hardware spends producing it. \scheme{} measures a property of the byte stream that neither family uses, at the cost of a compression pass. Since QAT as an accelerator reduces the latency and power consumption of compression significantly over software compression, the same benefit transfers to the detection power and latency as well.

We also build the detector on another built-in compression accelerator, Intel IAA, and compare it with the QAT-based detector to narrow down which accelerator qualities data-provenance detection requires (Section~\ref{subsec:other-accelerators}).

\noindent{\textbf{Contributions.}} This paper has the following contributions:

\begin{itemize}[leftmargin=*]

\item \textbf{Reverse engineering of Intel QAT and analysis of its sensitivity to AI-generated data.}
We present the first reverse engineering of a modern on-chip compression accelerator. It reveals the first known instance of an on-chip compression accelerator
whose amount of computation depends on byte-level repetition in
the data under compression.
It also provides the first observation that computation on AI-generated artifacts leaves
a measurably different microarchitectural footprint.
These results show that generated text, images, and code shift QAT
throughput and latency, whereas generated video shifts them less
(Section~\ref{subsec:modality-token-rate}).

\item
\textbf{Hardware basis of \scheme{}'s detection signal.}
We show that QAT's match finder turns an input's repetition structure into
time through three stages: a limited-capacity candidate table in which new
strings evict older repeats, a 48-byte comparator that confirms any shorter
match at one fixed cost, and an extender that completes longer matches at
extra cost.
Contrasts on identical inputs isolate these stages: CPU zlib, nvCOMP and
Intel IAA miss repeats only beyond a fixed distance, and zlib, which has no
comparator, pays for every matched byte.

\item \textbf{\scheme{}: A novel method for detecting AI-generated artifacts
and its power, accuracy, and latency advantages across modalities.}
\scheme{} consumes $1.2\times10^{-3}$\,J per check, compared with
\(24.9\,\mathrm{J}\) for Binoculars and \(3{,}258\,\mathrm{J}\) for DetectGPT,
and is 333\% faster in detection than Deepfake CLI.
Using the same hardware across all four modalities, \scheme{} obtains:
\circnum{1}~text: 100\% accuracy (0\% FPR, 0\% FNR) on DetectRL;
\circnum{2}~images: 98\% accuracy (98\% TPR, 2\% FPR) on Self-Synthesis'
BEGAN generator;
\circnum{3}~code: 70\% accuracy for Claude 3-written code from the STACAD
dataset (82\% TPR, 43\% FPR); and
\circnum{4}~video: 78\% accuracy (78\% TPR, 21\% FPR) on FETV/MSR-VTT.

\item \textbf{Robustness to paraphrasing and adaptive attacks.}
Against DIPPER paraphrasing, which rewrites AI text to sound human-written,
\scheme{} keeps 92\% accuracy on DetectRL (\(6.25\%\) FPR, \(9.38\%\) FNR)
and \(0.971\) AUROC,
versus \(0.462\) for DetectGPT.Against an adaptive attacker, who uses \scheme{}'s own scores to pick the most
human-looking of ten meaning-preserving rewrites, its AUROC on RAID drops only from \(1.00\) to \(0.96\), within the range of six software detectors under the same attack.

\end{itemize}

\noindent\textbf{Paper Organization.} Section~\ref{sec:attacks} establishes the hardware mechanisms behind the timing signal. Sections~\ref{sec:methodology} and~\ref{sec:qatscan-detect} present \scheme{}'s measurement and classifier procedures followed by detection results for text, code, images, and video, baseline comparisons, noise interference, and modality differences. Section~\ref{sec:related} discusses related work, and Section~\ref{sec:conclusion} concludes.

\section{\scheme{}}
\label{sec:attacks}\label{sec:lasso-import-obs2}\label{sec:component-inventory}%

\subsection{Overview}
\label{subsec:overview}
We find that QAT's timing has two stages.
First, a fixed-rate match finder with limited candidate capacity decides
which repeated byte sequences become back-references and which remain
literals.
Second, every request passes through a mandatory verification step, in
which an on-chip decompression engine rebuilds the input from the
compressed output; this engine spends about 1.05\,ns per literal and
2.0\,ns per back-reference, plus 0.16\,ns per matched byte beyond 48B.
QAT therefore acts as a \emph{microarchitectural footprint of
repeated-substring structure in the data itself}. Entropy and compression ratio cannot predict this because they only see ``how much redundancy is there in the data,'' whereas the hardware is sensitive to how the redundancy is packaged, which amplifies small structural differences between AI-generated and real data into measurable differences in throughput.

\subsection{Reverse Engineering}

\scheme{} rests on one claim: QAT's processing time depends on the
repetition structure of its input, and AI-generated media differ from
human-made media in that structure.
The detector is deliberately simple, a throughput measurement and a lightweight
classifier (Section~\ref{sec:methodology}), so its accuracy comes from the
timing signal rather than from the classifier. Trusting it therefore requires
knowing \emph{what} the timing measures and \emph{why}, which accuracy alone
cannot show. We answer three questions:
\textbf{\circnum{1}} which part of QAT produces the timing difference, and
why a CPU does not
(Sections~\ref{sec:qre-fixed-capacity}--\ref{sec:qre-evidence-6});
\textbf{\circnum{2}} why the signal survives paraphrasing
(Section~\ref{subsec:code-detection}); and
\textbf{\circnum{3}} what another accelerator would need to support the same
detection (Section~\ref{subsec:other-accelerators}).
We answer these with controlled inputs that differ only in how they repeat;
Section~\ref{sec:qatscan-detect} builds on the answers.

\begin{figure}[!htbp]
\centering
\includegraphics[width=0.8\columnwidth]{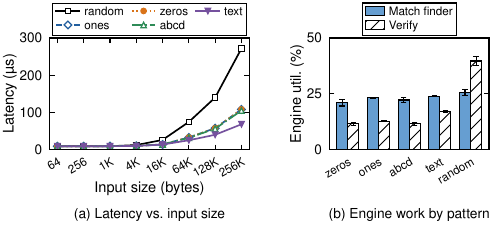}
\caption{(a) Random is slowest; (b) only verify work varies by pattern.}
\label{fig:lasso-import-latency}\label{fig:qre-engbypat}
\end{figure}

Figure~\ref{fig:detectrl_cpu_qat} shows that QAT separates human and AI text
under the same DEFLATE algorithm as the CPU.
However, real files differ in many properties at once, so this result cannot
show which property QAT responds to.
We therefore use synthetic inputs that differ mainly in how much they repeat:
zeros, ones, a repeating \texttt{ABCD} pattern, English text, and random bytes.
They span the range of \emph{entropy}, the average bits needed to encode each
byte,
from 0\,bit/B (all zeros) to 8\,bit/B (random).
We measure each pattern from 64\,B to 256\,KiB with static Huffman coding,
which uses DEFLATE's fixed code table~\cite{deutsch1996deflate} instead of
building one per request, and we record the QAT driver's 97 device-wide
performance and utilization counters.
Figure~\ref{fig:lasso-import-latency} plots (a) mean latency over five repeats
at compression levels L1--L4 and (b) compression- and decompression-engine
utilization (\texttt{util\_cpr0}, \texttt{util\_dcpr0}).

Figure~\ref{fig:lasso-import-latency}a exposes a counter-intuitive ordering.
Natural-language text has higher byte entropy (the average information per
byte: 4.3 versus 0.0--2.0\,bit/B) and compresses less ($12\times$ versus
$79$--$146\times$) than the repetitive zeros, ones, and ABCD inputs, yet it
completes fastest (74.2 versus 105--108.2\,\textmu s).
Random data, with the highest entropy and lowest compressibility, is slowest
(277.2\,\textmu s).
Engine utilization does not explain this ordering
(Figure~\ref{fig:lasso-import-latency}b): compression-engine utilization stays
at 21--27\% for every input, and decompression-engine utilization rises only
for random data (about 40\% versus 12--17\%).
Thus, latency follows neither how much information an input carries nor how
small it compresses.

Instead, latency tracks how an input is packaged into tokens, not what it contains. DEFLATE encodes its output as \emph{literals}, each emitting one byte, and \emph{back-references}, each a (length, distance) pair that copies
\emph{length} bytes from \emph{distance} bytes earlier.
Decoding QAT's output (Table~\ref{tab:char-backrefs}) shows three encodings:
random emits a literal for almost every byte (262{,}016 literals and 42
back-references); the repetitive inputs cover almost every byte with about
1{,}015 matches of about 258\,B; and text covers 99\% of its bytes with
6{,}902 matches of about 38\,B.
Latency falls in the reverse order of back-reference count, and
back-reference count depends on how the input's bytes repeat, not on how much
information they carry.

 \begin{center}\begin{minipage}{\columnwidth}\centering
{\footnotesize\tablesmaller\setlength{\tabcolsep}{2.5pt}%
\newcommand{\tOneIn}{\arrayrulecolor{black!22}\hline\arrayrulecolor{black}}%
\begin{tabular}{@{}lrrrrrrrr@{}}
\Xhline{0.9pt}
\textbf{pattern} & \textbf{entropy} & \textbf{comp.} & \textbf{back-} & \textbf{mean ref} & \textbf{literals} & \multicolumn{3}{c}{\textbf{latency (\textmu s)}} \\
\cline{7-9}
 & \textbf{(bit/B)} & \textbf{ratio} & \textbf{refs} & \textbf{len (B)} & & \textbf{pred.} & \textbf{meas.} & \textbf{$\boldsymbol{\Delta}$ (\%)} \\ \Xhline{0.6pt}
zeros  & $0.0$ & $81\times$  & $1{,}014$ & $258$ & $699$       & $103.7$ & $108.2$ & $-4.2$ \\ \tOneIn
ones   & $0.0$ & $79\times$  & $1{,}014$ & $258$ & $699$       & $103.7$ & $108.2$ & $-4.2$ \\ \tOneIn
abcd   & $2.0$ & $146\times$ & $1{,}017$ & $258$ & $8$         & $103.2$ & $105.0$ & $-1.7$ \\ \tOneIn
text   & $4.3$ & $12\times$  & $6{,}902$ & $38$  & $1{,}430$   & $77.7$  & $74.2$  & $+4.7$ \\ \tOneIn
random & $8.0$ & $0.9\times$ & $42$      & $3$   & $262{,}016$ & $273.0$ & $277.2$ & $-1.5$ \\ \Xhline{0.9pt}
\end{tabular}}
\vspace{2pt}\captionof{table}{Token counts, not entropy, explain QAT latency at 256\,KB.}\label{tab:char-backrefs}
\end{minipage}\end{center}

\subsection{Inside the Match Finder: Table, Comparator, and Extender}
\label{sec:qre-fixed-capacity}

\begin{figure}[!htbp]
\centering
\resizebox{\columnwidth}{!}{%
\begin{tikzpicture}[font=\fontsize{7}{8}\selectfont,
  st/.style={draw=black,fill=white,rounded corners=0pt,align=center,
    minimum height=7.5mm,inner sep=1.5pt,line width=.5pt},
  hi/.style={st,line width=.9pt},
  ar/.style={-{Stealth[length=3pt]},line width=.55pt,draw=ink2},
  pr/.style={font=\fontsize{6}{7}\selectfont\itshape,text=ink2,align=center}]
\node[st] (h) {Input\\hash};
\node[st,right=3mm of h] (t) {Candidate table\\(limited capacity)};
\node[hi,right=3mm of t] (c) {Comparator\\($\le$48\,B, flat cost)};
\node[hi,right=3mm of c] (e) {Extender\\($>$48\,B, waits on history)};
\node[st,right=3mm of e] (tok) {Literals,\\back-refs};
\node[st,right=3mm of tok] (v) {Verify:\\decompress};
\draw[ar] (h)--(t); \draw[ar] (t)--(c); \draw[ar] (c)--(e);
\draw[ar] (e)--(tok); \draw[ar] (tok)--(v);
\draw[ar,dashed] (t.south) -- ++(0,-4.2mm) -| (tok.south);
\node[pr,below=.3mm of t] {new candidates overwrite old};
\node[pr,below=4.4mm of e.south,anchor=north] {miss $\rightarrow$ literal};
\node[pr,above=.3mm of v] {1 cycle per symbol};
\end{tikzpicture}}
\caption{Inferred QAT Pipeline. A candidate that survives the table is confirmed by a 48-byte comparator. Both compression and verify engines overlap in time; the verify engine retires one Huffman symbol per cycle.}
\label{fig:qre-model}
\end{figure}
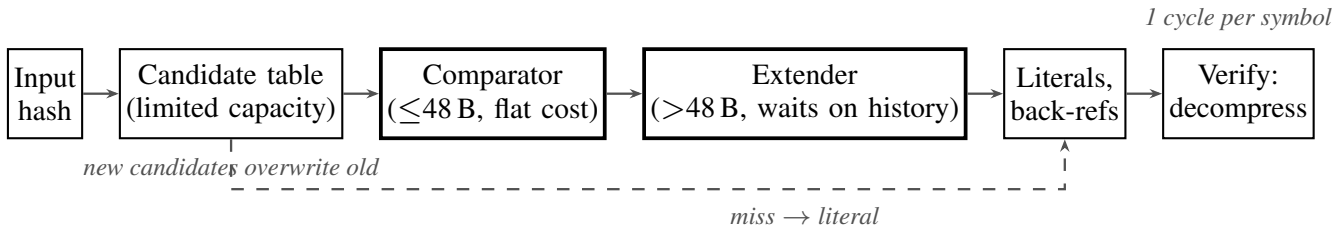

QAT is a black-box accelerator: software reaches it only through an API, and
its documentation specifies that interface but not the internals of the
match search.
The only timing visible through this API is the latency of a whole request,
which aggregates the work spent on every byte of the input.
The driver's telemetry counters report device activity, such as how busy each
engine is, but only as totals over the whole device.
Thus, neither timing nor telemetry can reveal how the engine handled any
particular position in the input.
The compressed stream, in contrast, records the match finder's decision for
every byte: whether it was emitted as a literal or copied by a
back-reference, and if copied, from how far back and as part of how long a
match.
Therefore, we reverse engineer the match search from the compressed streams
that each engine produces.
Figure~\ref{fig:qre-model} summarizes the design that the following
experiments infer, and labels each stage with the figure that probes it.

To separate QAT's design from the DEFLATE format, we give identical inputs to
four implementations: QAT 401xx; Intel IAA~\cite{intel_qpl}, a second on-die
compression accelerator, through the hardware path of the Intel Query
Processing Library; the Deflate codec of nvCOMP
5.3.0~\cite{nvidia_nvcomp} on an A100 GPU; and zlib 1.2.13~\cite{zlib} at level~1 on
the host CPU.
QAT alone runs every request as compress-and-verify, decompressing its own
output to check it before completing, and this pass cannot be disabled.
For each implementation we measure compressed size and request latency, and
decode each output token by token to see which repeats became matches.

\noindent\textbf{How far back can a repeat be matched?}
Each input contains a random 64-byte pattern twice, with the second copy
starting $D$ bytes after the first and random filler between them.
Thus, the first copy is the only earlier position that can supply the
repeat, and an engine can encode the second copy as a back-reference only if
its match finder still holds the first copy when the repeat arrives.
We verify that each stream decompresses to the original input and decode it
token by token.
A trial is a \emph{hit} only if the token at the first byte of the second copy
is a back-reference with distance exactly $D$; any other outcome is a
\emph{miss}.
For each $D$, we run 1{,}000 trials with new random data.
All engines compress identical inputs, each in a single request.
Therefore, differences in miss rate come from how each engine finds matches.

\begin{figure}
    \centering
    \includegraphics[width=\linewidth]{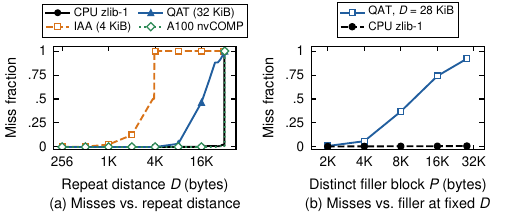}
    \caption{Match finder: (a) misses within the history window, (b) rise with distinct filler}
    \label{fig:qre-mf}
\end{figure}

CPU zlib-1 and A100 nvCOMP miss almost no repeats up to 32\,KiB and every
repeat beyond it; IAA misses every repeat beyond 4\,KiB
(Figure~\ref{fig:qre-mf}a).
This step is what a distance limit produces.
QAT shows no such step: its miss rate rises from 0.03 at 8\,KiB to 0.46 at
16\,KiB, a distance that zlib and nvCOMP reach, and to 0.96 near 31\,KiB.
Thus, distance alone does not explain QAT's misses.
We hypothesize that QAT's match finder is a table of limited capacity: each
filler position inserts a candidate, and once enough distinct strings compete
for the first copy's entry, they evict it.

\noindent\textbf{Keeping distance constant.}
To test this hypothesis, we fix $D = 28$\,KiB, inside zlib's 32{,}506-byte
maximum match distance, and vary only how many distinct strings lie between
the two copies.
We build filler by repeating a random block of $P$ bytes, so the filler holds
about $P$ distinct strings; $P = D$ makes it fully random.
If distance causes the misses, the miss rate should stay high for every $P$;
if eviction causes them, it should start near zero and rise with $P$.

QAT's miss rate stays at or below 0.01 for $P \le 2{,}048$, even though every
repeat is 28\,KiB away (Figure~\ref{fig:qre-mf}b).
It then rises to 0.06 at $P = 4{,}096$, 0.37 at 8{,}192, 0.74 at 16{,}384, and
0.92 for fully random filler.
The same $P$ gives nearly the same miss rate at $D = 16$ and 24\,KiB (for
example, 0.065 at $P = 4{,}096$), and CPU zlib stays at or below 0.01 for every
$P$, so the filler itself does not cause misses.
Therefore, QAT's match finder has limited capacity, and new distinct strings
evict older candidates.
With random filler, every additional byte adds a distinct string, so a more
distant repeat faces more evicting strings; thus, distance also raises QAT's
miss rate (Figure~\ref{fig:qre-mf}a), through eviction rather than a hard
distance limit.

\begin{figure}
    \centering
    \includegraphics[width=0.9\linewidth]{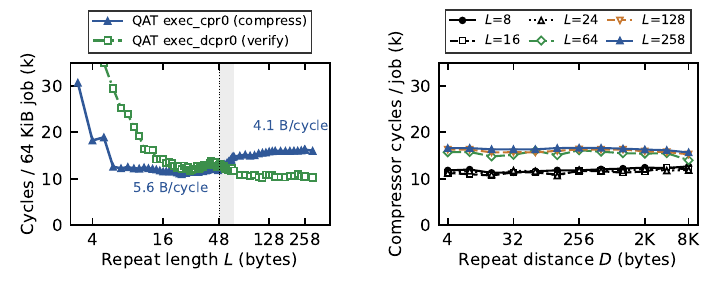}
    \caption{Compression-engine work: (a) flat to 47\,B, then rising from 48\,B as
matches reach the extender; (b) independent of distance at every match length.}
    \label{fig:qre-mf_fig8}
\end{figure}

\noindent\textbf{How long a match can the comparator confirm?}
A match finder confirms a candidate by comparing the input against the
history; the comparator's \emph{width} is the number of bytes it compares in
one fixed-cost operation, and longer matches need another mechanism.
To measure the width, each input repeats a unit of $L$ bytes copied from
$D = 2{,}000$ bytes back, followed by one random breaker byte that ends the
match.
We sweep $L$ from 3 to 128 in 1-byte steps, plus selected lengths up to 300.
Thus, distance and input size stay fixed, and only how the bytes are packaged
into matches changes.
Request latency cannot isolate the comparator, because the compression and
verify engines overlap and latency reports only the longer of the two.
We therefore report per-engine busy cycles per job (\texttt{exec\_cpr0} for
compression, \texttt{exec\_dcpr0} for verify) under a sustained load of
64\,KiB jobs.
If the comparator is $W$ bytes wide, cycles per job should be flat for
$L \le W$ and rise beyond it; if match length never matters to the
matcher, they should be flat throughout.

The compression slice spends 11.0--12.2\,k cycles per job for every $L$ from
6 to 47 (Figure~\ref{fig:qre-mf_fig8}a), about 5.6\,B/cycle for both 8- and
47-byte matches.
At $L = 48$ it rises to 12.5\,k and keeps rising before leveling at 16.0--16.3\,k up to 258 (about 4.1\,B/cycle).
The verify slice falls over the same range, from 50\,k cycles at $L = 3$ to
about 10\,k above 64, as it decodes fewer symbols; thus, the rise is not due to work
shifting between engines.
On the same kind of input, software zlib level~1, which has no comparator
and checks a match one byte at a time, pays 0.61\,ns for every matched byte
with no change at 48\,B, while QAT pays nothing up to 47\,B
(Figure~\ref{fig:qre-cpu}).
Thus, the comparator confirms matches of up to 47 bytes at one fixed cost,
and 48 bytes is the first length that costs more.

\noindent\textbf{What completes a match longer than 48 bytes?}
If the same comparator simply ran longer for long matches, the extra cost
would appear as execution and would not depend on where the earlier copy
lies.
We test eleven lengths at twelve distances from 4 to 8{,}192 bytes in Figure~\ref{fig:qre-mf_fig8}b.
Matches of 8, 16, and 24 bytes cost the same at every distance: the
compression counter stays within $\pm$5\% of 11.5\,k and the latency excess
within $\pm$2\,$\mu$s.
For 258-byte matches, the counter is again flat (16.3--16.6\,k), but the
latency excess varies with distance: 49--51\,$\mu$s for $D \le 512$. 
Relative to short matches, the counter rises by about 18\,$\mu$s of cycles
while latency rises by about 41\,$\mu$s, so more than half of the long-match
time is spent with the slice not executing.
Thus, a separate stage, which we call the \emph{extender}, completes matches
beyond the comparator and waits on history for a distance-dependent time.
Short matches never enter it, so distance adds no measurable cost to them.

\subsection{Evidence for Per-Token Latency Cost}
\label{sec:qre-evidence-6}

\begin{figure}[t]
\centering
\includegraphics[width=0.7\columnwidth]{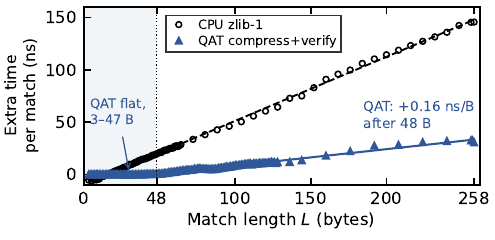}
\caption{Extra time per match relative to $L = 16$. QAT is flat up to
47\,B and then rises by 0.16\,ns/B; software zlib level~1 rises by
0.61\,ns/B at every length.}
\label{fig:qre-cpu}
\end{figure}

Section~\ref{sec:qre-fixed-capacity} established which repeats the match
finder turns into back-references; this subsection measures, empirically,
how much request latency each resulting token adds.
We do not derive these costs from the circuit; we measure them on inputs
whose token streams we control and verify by decoding the compressed output.
Latency is flat at 4.5\,\textmu s for all inputs from 64\,B to 256\,B.
This is a fixed per-request overhead rather than data processing, and we
subtract it from every measurement below.

\noindent\textbf{Cost of a literal.}
Random input is encoded almost entirely as literals (262{,}016 of
262{,}058 tokens).
Subtracting the overhead from the latency of a 256\,KiB random input and
dividing by the literal count gives 1.05\,ns per literal.
Two independent measurements agree: the input-size sweep gives 1.0\,ns per
byte for random data, and decompressing a hand-built all-literal stream
gives 1.04\,ns per literal.
Thus, empirically, each literal adds about 1.05\,ns.

\noindent\textbf{Cost of a match.} We measure it as the latency difference
between two inputs that differ only in their number of matches. Hand-built
streams that all decompress to the same 256\,KiB with the same 64 literals
use matches of 3 to 10 bytes, short enough that decoding, not writing the
output, sets the time. The 3-byte stream (87{,}360 matches) takes
122.6\,\textmu s longer than the 10-byte stream (26{,}208 matches), or
2.00\,ns per match; every adjacent pair from 3 to 10 bytes gives
1.96--2.05\,ns, and a repeated run gives 2.01\,ns. Because the output and
the literal count are the same in every stream, fixed and per-literal costs
cancel. Thus each match adds about 2.0\,ns. Longer matches add more: the
time compression adds beyond decoding, per match, stays flat up to 47\,B
and then rises by 0.16\,ns per byte (Figure~\ref{fig:qre-cpu}). 

\noindent\textbf{Empirical latency model.}
The match finder's output is described by three quantities: the literal
count $n_{\mathrm{lit}}$, the match count $n_{\mathrm{match}}$, and the
match lengths $\ell_i$.
Combining the measured costs gives an empirical model of request latency:
\begin{equation}
\begin{aligned}
\widehat{T}_{\mathrm{QAT}} &=\quad a + b\biggl[1.05\,n_{\mathrm{lit}} + 2.0\,n_{\mathrm{match}}\\
&\qquad{}+0.16\sum_{i=1}^{n_{\mathrm{match}}}\max(0,\ell_i-48)\biggr].
\end{aligned}
\label{eq:qre-costmodel}
\end{equation}
Here, $\widehat{T}_{\mathrm{QAT}}$ is the predicted request latency, and
$a$ and $b$ are an offset and scale fitted to the measurements.
The model is empirical: it summarizes the measured costs and makes no claim
about how QAT is built.
It contains no term for byte values or hash-table occupancy.
With $a$ and $b$ fitted, it predicts each pattern's latency within
\textbf{5}\% and reproduces their order, from text (fastest) to random
(slowest) (Table~\ref{tab:char-backrefs}).

\noindent\textbf{What the reverse engineering means for provenance.}

Each stage of the inferred path (Figure~\ref{fig:qre-model}) shapes how an
input's repetition structure becomes time. First, the candidate table
decides which repeats count: a repeat is found only while its earlier copy
survives the distinct strings inserted after it (Figure~\ref{fig:qre-mf}),
so QAT measures local rather than document-wide repetition. On our 64\,KB
samples, GPT-2 text has a larger fraction of distinct 3-byte strings than
human text (0.10 vs.\ 0.08), which leaves more bytes with no earlier copy
and fills the table faster; both turn bytes into literals. Second, the
comparator charges the same for any match up to 47\,B
(Figures~\ref{fig:qre-mf_fig8}a and~\ref{fig:qre-cpu}); with mean match
lengths of 7.15\,B (human) and 6.65\,B (GPT-2), nearly every text match
falls in this range, so a longer human match covers more bytes at no extra
cost. Third, the extender, which completes matches beyond 48\,B, barely
affects text but dominates the zeros, ones, and ABCD patterns, whose
258-byte matches make them slower than text despite far better compression
(Figure~\ref{fig:lasso-import-latency}). Fourth, the per-token costs (Section~\ref{sec:qre-evidence-6}) convert token
counts into time: with lower byte coverage (87.3\% vs.\ 90.3\%), a GPT-2 block
carries about 1{,}960 more literals and 320 more matches than a human block,
predicting 2.7\,\textmu s of added latency against a measured 1.8\,\textmu s
(29{,}500 vs.\ 28{,}000 requests/s for the block itself). Therefore, on text, the block's QAT latency reports how many bytes the match finder could not cover with a nearby exact repeat; GPT-2 text, which repeats itself less locally, leaves more such bytes and runs slower. The detection feature, the monitor process's throughput, rises when the block compresses faster; the classifier learns the sign of the human--AI gap.
Compression ratio counts the same literals but prices them in bits, not
processing time, so blocks with equal ratio can still differ in QAT time.
This is why Lasso retains 66\% accuracy on paraphrased text at fixed
compression ratio, and why Section~\ref{subsec:modality-token-rate}
summarizes every modality by its token rate.

\subsection{What Another Accelerator Needs}
\label{subsec:other-accelerators}
The comparisons above identify three properties that let an accelerator's
timing reveal provenance.
First, its cost must follow tokens rather than bytes: QAT charges any match of
up to 47\,B the same, so its latency tracks how many literals and matches an
input needs, whereas zlib pays for every matched byte
(Figure~\ref{fig:qre-cpu}) and does not separate human from AI text
(Figure~\ref{fig:prelim-silicon}).
Second, which repeats it finds should depend on the input's local repetition:
QAT's misses grow with the number of distinct strings between a repeat and its
earlier copy, whereas zlib, nvCOMP and IAA miss repeats only beyond a fixed
distance (Figure~\ref{fig:qre-mf}); IAA, limited to 4\,KiB, supports a weaker
detector (Figure~\ref{fig:raid-transfer}).
Third, its timing must be stable enough to resolve a gap of about 5\% per
block: across 6{,}000 repetitions, the interquartile range of QAT's request
latency is at most 0.5\% of its median.

\subsection{Why Should This Signal Be Robust Against Evasion?}
\label{subsec:evasion-signal}
Figure~\ref{fig:roc-sweep} reports \scheme{}'s strong detection accuracy after DIPPER~\cite{krishna2023dipper} paraphrasing (Section~\ref{sec:intro}). DIPPER substantially changes wording and sentence structure, disrupting the signature other detectors use, yet QAT throughput remains a useful detection signal. Reverse engineering (Section~\ref{sec:attacks}) explains causation of detection through matching, which runs at an approximately fixed rate, determines the emitted literals and back-references caused by QAT's pipeline.

\begin{figure}[!htbp]
  \centering
  \includegraphics[width=0.8\columnwidth]{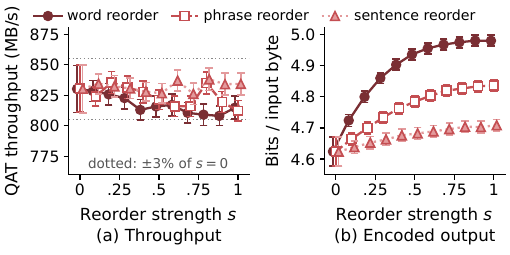}
  \caption{Reordering raises bits/byte; throughput stays within $\pm$3\%.}
  \label{fig:paraphrase-reorder-throughput}
\end{figure}

\looseness=-1 \noindent\textbf{Different matches can have similar processing costs.}
The per-token measurements (Section~\ref{sec:qre-evidence-6}) and the latency model (Equation~\ref{eq:qre-costmodel}) explain how this occurs. 
\looseness=-1 A DIPPER paraphrase can change \emph{The storm destroyed the gate at dawn, leaving the road open} to \emph{At dawn, the gate was destroyed by the storm, leaving the road open}. This reordering can break the long match to \emph{the storm destroyed the gate} while retaining shorter repeated substrings such as \emph{at dawn} and \emph{leaving the road open}. The added setup of the extra short matches can thus be offset by less long-match expansion. In a controlled reordering sweep on real text, Figure~\ref{fig:paraphrase-reorder-throughput} shows that encoded output (bits per byte) grows with reordering strength while measured QAT engine throughput stays within a narrow band.

\looseness=-1 \noindent\textbf{Connection to DIPPER.}
DIPPER~\cite{krishna2023dipper} is a paraphrasing method that evades and fools many state of the art AI detectors. 
It conditions on a passage and optional surrounding context, applies lexical- and order-diversity controls, and generates a rewrite that aims to retain meaning. 
Its order-diversity control corresponds to our controlled reordering sweep, and its lexical changes can remove repeated strings while retaining or creating others.
\looseness=-1 Together, the reordering sweep identifies a mechanism by which linguistic cues can change substantially while the QAT signal changes less, and the DIPPER~\cite{krishna2023dipper} evaluation establishes that sufficient throughput separation remains for detection.

\subsection{Detection Steps}
\label{subsec:classifier}
\scheme{} trains a classifier on these measurements for a reference
dataset of artifacts whose provenance is known.
For text and software code, the classifier receives one scalar: throughput.
\scheme{} uses an $L_2$-regularized logistic regression on standardized features.
Each reference artifact $i$ has a known label $y_i\in\{0,1\}$ and a scalar feature $x_i=\bar t_i$, the monitor process's throughput during the single compression of artifact $i$. Training is deterministic: 
\begin{enumerate}
  \item \textbf{Standardize:} map each throughput $x_i$ to
        $z_i=(x_i-\mu)/s$, using the reference mean $\mu$ and standard
        deviation $s$, so the fit does not depend on units or scale.
  \item \textbf{Fit:} learn one weight $w$ and bias $b$ by minimizing the
        $L_2$-regularized logistic loss of $p_i=\sigma(wz_i+b)$.
  \item \textbf{Decide:} an unseen artifact with throughput $x$ scores
        $p=\sigma\!\big(w(x-\mu)/s+b\big)$ and is labeled AI-generated
        when $p\ge0.5$. This is equivalent to a threshold
        $t^\star=\mu-s\,b/w$ on throughput: $x\ge t^\star$ if $w>0$, and
        $x\le t^\star$ if $w<0$. The sign of $w$ is learned, so the rule
        does not assume which class has the higher throughput. We report
        the AUROC of $p$ on held-out data. AUROC is the probability that
        a random AI-generated sample scores higher than a random human
        one ($1.0$ is a perfect ordering, $0.5$ is chance), and it does
        not depend on $t^\star$.
\end{enumerate}

If throughput alone does not separate the classes, the same procedure runs on an
extended feature vector, making $w$ a vector: the throughput of each of several fixed-size 64\,KB blocks from different positions in the file (we call each block a \emph{crop}), with request latency; steps (1)--(3) are unchanged. The method uses
no neural network, no access to the generative model, and no pixel- or
token-level analysis. 

\vspace{-1em}
\subsection{Multimodality Detection}
\label{subsec:modality-token-rate}

QAT runs the same pipeline on text, images, video, and code, and its work
is the weighted token cost of the stream it produces
(Equation~\ref{eq:qre-costmodel}). We therefore summarize every input,
whatever its modality, by its token rate (QAT tokens per KB). The rate
describes match structure, not work: literals, matches, and match bytes
beyond 48 cost differently, and text emits five times the tokens of
all-zeros yet completes faster (Table~\ref{tab:char-backrefs}). Two
properties of the rate differ across modalities. The first is its natural
variation between inputs, which sets how large a human--AI shift must be to
stand out. At a common 16\,KB unit, the coefficient of variation is
$28.6\%$ for text, $33.4\%$ for images, and $38.4\%$ for single video
frames. Images have the largest SD ($191$ vs.\ $37$ and $99$ tokens/KB)
but also the highest mean. Larger units reduce the variation, to $17.8\%$
for 64\,KB text blocks and $23.2\%$ for per-video medians. The second is
how far generated content shifts the rate: clearly for text and images,
whereas real and generated video frames overlap.

\begin{figure}[!t]
  \centering
  \includegraphics[width=0.7\linewidth]{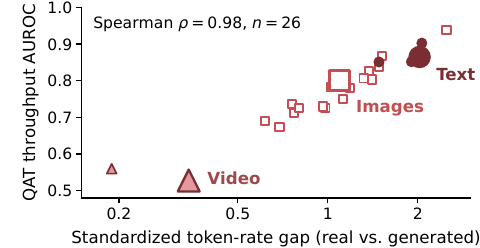}
  \caption{Token-rate gap predicts AUROC in every modality.}
  \label{fig:modality}
\end{figure}

\looseness=-1 Across $26$ dataset/generator settings
(Figure~\ref{fig:modality}), the AUROC of QAT throughput rises with the
token-rate gap divided by the within-class standard deviation (Spearman
$\rho=0.98$), and all three modalities fall on the same relation. The text
shift is about twice the natural variation (AUROC $0.86$), and the image
shift about equal to it ($0.80$). For video, whose frames overlap, the
detector aggregates many frames and crops ($0.53$). Therefore, one
measurable input property, the standardized token-rate gap, predicts
\scheme{}'s accuracy in every modality.

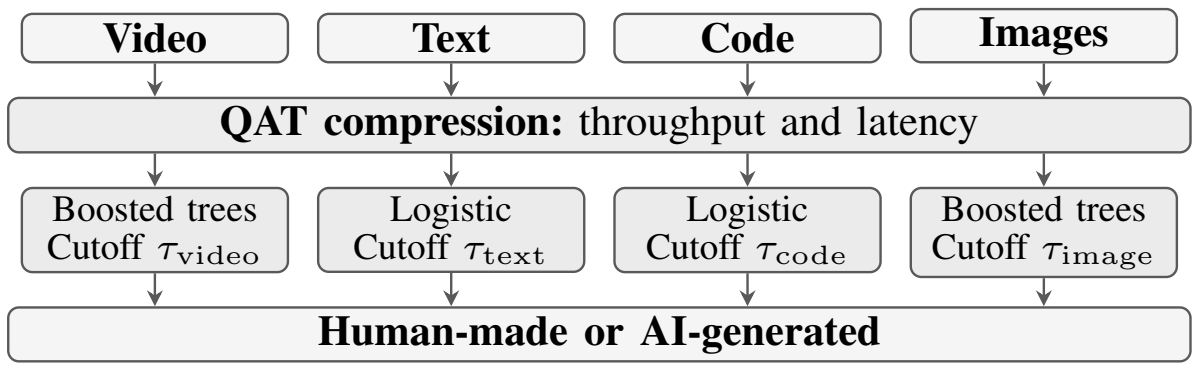
\begin{figure}[!ht]
\centering
\begin{tikzpicture}[
  every node/.style={font=\footnotesize,align=center},
  block/.style={draw=black!65,rounded corners=2pt,line width=.45pt,inner xsep=3pt,inner ysep=1.5pt},
  media/.style={block,fill=black!4,minimum width=1.76cm,minimum height=.32cm},
  cutoff/.style={block,fill=LRED!12,minimum width=1.76cm,minimum height=.4cm,font=\scriptsize},
  flow/.style={-{Stealth[length=3pt,width=3pt]},line width=.45pt,draw=black!65}
]
\node[block,fill=black!7,minimum width=7.82cm,minimum height=.32cm]
  (qat-compression) at (0,0) {\textbf{QAT compression:} throughput and latency};
\foreach \xpos/\kind/\mediaName/\modelName in {
  -2.94/video/Video/{Boosted trees}, -.98/text/Text/Logistic,
  .98/code/Code/Logistic, 2.94/image/Images/{Boosted trees}
}{
  \node[media,anchor=south] (input-\kind) at ([yshift=.22cm]\xpos,0 |- qat-compression.north) {\textbf{\mediaName}};
  \node[cutoff,anchor=north] (score-\kind) at ([yshift=-.22cm]\xpos,0 |- qat-compression.south) {\modelName\\[-1pt] Cutoff $\tau_{\mathrm{\kind}}$};
}
\node[block,fill=black!4,minimum width=7.82cm,minimum height=.32cm,anchor=north]
  (provenance) at ([yshift=-.22cm]qat-compression |- score-text.south) {\textbf{Human-made or AI-generated}};
\foreach \kind in {video,text,code,image}{
  \draw[flow] (input-\kind.south) -- (input-\kind.south |- qat-compression.north);
  \draw[flow] (score-\kind.north |- qat-compression.south) -- (score-\kind.north);
  \draw[flow] (score-\kind.south) -- (score-\kind.south |- provenance.north);
}
\end{tikzpicture}
\caption{\scheme{} calibrates a separate decision threshold for each modality.}
\Description{Video, text, code, and images enter the same QAT compression pipeline. Separate classifiers and thresholds use the resulting timing features to classify each input as human-made or AI-generated.}
\label{fig:modality-thresholds}
\end{figure}

\looseness=-1 \scheme{} selects the observation unit, feature aggregation, and classifier by modality, with a separately calibrated decision threshold (Figure~\ref{fig:modality-thresholds}). As new generators appear, we can update each classifier and threshold with labeled human-made samples and outputs from recent generators, then evaluate robustness on held-out generators.

\begin{table*}[!t]
\centering\footnotesize

\definecolor{tTwoLasso}{HTML}{F6E3E4}%
\newcommand{\tTwoIn}{\arrayrulecolor{black!22}\hline\arrayrulecolor{black}}%
\newcommand{\tTwoGrp}[2]{\multirow{#1}{*}{\rotatebox[origin=c]{90}{\scriptsize\textbf{#2}}}}%
\setlength{\tabcolsep}{3pt}\renewcommand{\arraystretch}{1.0}
\begin{tabularx}{\textwidth}{@{}c@{\hspace{3pt}}l*{5}{>{\centering\arraybackslash}X}>{\columncolor{tTwoLasso}\centering\arraybackslash}X@{}}
\Xhline{0.9pt}
 & \textbf{Metric} & \textbf{Log-Rank} & \textbf{Fast-DetectGPT} & \textbf{Binoculars} & \textbf{DetectGPT} & \textbf{DNA-GPT} & \textbf{\scheme{}}\\ \Xhline{0.6pt}
\tTwoGrp{9}{Detection quality} & AUROC: Original / No evasion & 0.883 & 0.874 & 0.943 & 0.786 & 0.608 & \textbf{1.000}\\ \tTwoIn
 & AUROC: Paraphrase avg. & 0.677 & 0.732 & 0.808 & 0.462 & 0.742 & \textbf{0.971}\\ \tTwoIn
 & AUROC reduction after paraphrasing (pp) & 20.6 & 14.2 & 13.5 & 32.4 & $-13.4$ & \textbf{2.9}\\ \tTwoIn
 & AUROC: Perturbation avg. & 0.394 & 0.459 & 0.769 & 0.132 & 0.674 & \textbf{0.849}\\ \tTwoIn
 & AUROC reduction after perturbation (pp) & 48.9 & 41.5 & 17.4 & 65.4 & $-6.6$ & \textbf{15.1}\\ \tTwoIn
 & Accuracy: Original / No evasion & 0.807 & 0.840 & 0.903 & 0.762 & 0.600 & \textbf{1.000}\\ \tTwoIn
 & Accuracy: Paraphrase avg & 0.662 & 0.711 & 0.781 & 0.554 & 0.637 & \textbf{0.921}\\ \tTwoIn
 & \textbf{Accuracy: overall avg} & \textbf{0.589} & \textbf{0.654} & \textbf{0.773} & \textbf{0.529} & \textbf{0.620} & \textbf{0.818}\\ \tTwoIn
 & \makecell[l]{Confusion matrix\\$[TP/FN/FP/TN]$}
  & \makecell{[34.3/65.7\\16.5/83.5]}
  & \makecell{[38.4/61.6\\7.6/92.4]}
  & \makecell{[58.9/41.1\\4.3/95.7]}
  & \makecell{[24.6/75.4\\18.9/81.1]}
  & \makecell{[80.7/19.3\\56.8/43.2]}
  & \makecell{\textbf{[71.1/28.9}\\\textbf{7.5/92.5]}}\\ \Xhline{0.6pt}
\tTwoGrp{3}{Gain} & \scheme{} AUROC gain: Original (pp) & +11.7 & +12.6 & +5.7 & +21.4 & +39.2 & ---\\ \tTwoIn
 & \scheme{} AUROC gain: Paraphrase (pp) & +29.4 & +23.9 & +16.3 & +50.9 & +22.9 & ---\\ \tTwoIn
 & \scheme{} AUROC gain: Perturbation (pp) & +45.5 & +39.0 & +8.0 & +71.7 & +17.5 & ---\\ \Xhline{0.6pt}
\tTwoGrp{4}{Cost} & Detection latency & 70 ms & 71 ms & 234 ms & 17.4 s & 3.75 s & \textbf{197 $\boldsymbol{\mu}$s}\\ \tTwoIn
 & Energy per check (J) & 16.3 & 16.3 & 24.9 & 3,258 & 369 & $\boldsymbol{1.2\times10^{-3}}$\\ \tTwoIn
 & \scheme{} speedup & 356$\times$ & 359$\times$ & 1,188$\times$ & 88,231$\times$ & 19,014$\times$ & ---\\ \tTwoIn
 & \scheme{} energy reduction (\%) & 99\% & 99\% & 99\% & 99\% & 99\% & ---\\ \Xhline{0.6pt}
\tTwoGrp{5}{Deployment} & Detector complexity & High & High & High & High & High & \textbf{Simple}\\ \tTwoIn
 & Robustness to semantic-preserving paraphrasing & Low & Medium & Medium & Very low / unreliable & Medium & \textbf{High}\\ \tTwoIn
 & Real-time latency target ($\leq100$ ms) & Yes & Yes & No & No & No & \textbf{Yes}\\ \tTwoIn
 & Edge-device average power ($\leq$5--6 W at 10 checks/s) & No & No & No & No & No & \textbf{Yes}\\ \tTwoIn
 & Sustainability (higher = lower energy per check) & Medium & Medium & Medium & Low & Low & \textbf{High}\\ \Xhline{0.9pt}
\end{tabularx}
\vspace{2pt}\caption{DetectRL text detection with DIPPER paraphrases; cost is per check: one 1.28\,KB document (software) vs. one 64\,KB block (\scheme{}).}%

\label{tab:detectrl-baseline-cost}
\end{table*}

\section{Methodology}
\label{sec:methodology}

\subsection{Platform and Measurement}
\label{subsec:platform}
Unless stated otherwise, all measurements use the default configuration.
The accelerator is the Intel QAT 401xx in a Xeon Gold 5420+ with static
Huffman coding, which uses DEFLATE's fixed code table rather than a table built for each request.
Each sample is a 64\,KB block, the size of one QAT request; for images, the
block is the first 64\,KB of the image file.
Each block is compressed once while a monitor process on another core compresses on the same QAT device; throughput is the monitor's rate (jobs/s), which rises when the block compresses faster.
Experiments that depart from this configuration state the difference where
they are introduced.
We estimate edge suitability from the energy measured on the server's QAT~401xx, the same accelerator family Intel integrates in its edge processors~\cite{intel-xeon-d}.

\noindent\textbf{Multi-workload conditions.} We evaluate Lasso in a multi-workload setting where a co-tenant load generator drives the \emph{same} QAT device as the detector with
$N$ compression threads, each of which submits a new request the moment its previous one completes, with no pause between requests (unthrottled); $N$ is the \emph{noise level}, swept over
$\{0,1,2,4,6,8,10,12,14\}$ ($N{=}0$ adds no load beyond the monitor process, and $N{=}14$ is the heaviest load;
engine utilization is pinned at $100\%$ from $N{=}4$, Fig.~\ref{fig:noise}(d)). The detector is trained once on a quiet machine ($N{=}0$): it
learns only the orientation of the real-vs-generated throughput difference, which is \emph{frozen}, that is, its threshold and orientation are held fixed rather than retrained, and evaluated unchanged at every noise level on disjoint
test content. Throughput is the only feature; compressed size is not used.

\subsection{Benchmarks}
\label{subsec:benchmarks}
\noindent\textbf{Text and code.} WikiText-103~\cite{merity2016pointer} tests matched content ($80/41$ train/test articles, article-disjoint); Beemo~\cite{beemo} tests AI editing with matched content ($1{,}516/606$ pairs, prompt-disjoint); FAIDSet~\cite{ta2025faid} tests academic writing (human $3{,}884/674$, AI $1{,}710$--$3{,}372/317$ per family); and STACAD~\cite{stacad} tests function-preserving code rewrites ($50/50$, source-file-disjoint).

\noindent\textbf{Generator and domain transfer.} In RAID~\cite{raid}, human and AI texts share titles, equalized lengths, and normalization.
Documents are split equally between training and testing before packing. The cross-generator evaluation trains on one source generator and tests on a target generator. Training does not pool generators, and the target generator supplies no training examples.
A detector may encounter text from a generator or subject area absent during training. Training on one generator and testing on another tests whether \scheme{}'s hardware signal transfers to an unseen generator.
Holding out each domain tests transfer across subject matter.

\noindent\textbf{Rewriting and evasion.} DetectRL~\cite{wu2024detectrl} covers four domains (arXiv abstracts, XSum news, Writing Prompts stories, and Yelp reviews) and four generators (GPT-3.5, PaLM-2, Claude-instant, and Llama-2-70B).
The training set contains $10{,}192$ human-written and $10{,}192$ AI-generated texts ($20{,}384$ total); the test set contains $1{,}008$ texts per class ($2{,}016$ total). The threshold is fitted on unmodified training text and applied unchanged to the transformed test text. DIPPER~\cite{krishna2023dipper} is a paraphrasing model with two controls: \emph{lexical diversity}, and \emph{order diversity}, each on a 0--100 scale. We use the $40/40$ setting, which rewrites multi-sentence passages while preserving meaning.

\noindent\textbf{Images and video.} We evaluate \scheme{} on photographs, generated and edited faces, diffusion images, document images, and video. Image experiments use GenImage~\cite{zhu2023genimage} and Self-Synthesis~\cite{tan2024freqnet} used by Zhou et al.~\cite{zhou2026beyondsemantic}, DRCT-2M~\cite{chen2024drct} for matched AI/Human images, CLIC\slash DiffusionDB~\cite{wang2022diffusiondb}, and SRD/GPT-4o receipts~\cite{expressexpense2019srd,zhang2026gpt4oreceipt}. Video experiments examine face manipulation~\cite{dufour2019dfd,roessler2019faceforensicspp} and matched-codec generation~\cite{liu2023fetv, xu2016msrvtt}.
GenImage tests shared visual classes; Self-Synthesis tests face generation and editing (identity-disjoint; $2{,}000$ per class per generator); DRCT-2M tests generated images with matched content (grouped by source image); CLIC\slash DiffusionDB tests matched resolution and PNG format ($32$ per class total); and SRD/GPT-4o tests receipt forgery ($200$ per class total). All image benchmarks use 5-fold cross-validation. FaceForensics++ DFD tests face swaps under an observation budget (train/test: $19/9$ actors, $478/232$ human/AI videos), and FETV/MSR-VTT tests matched encoding and frame budget (human $308/310$, AI $262/262$).

\subsection{Baselines}
\label{subsec:baselines}
Table~\ref{tab:detectrl-baseline-cost} compares \scheme{} with five detectors, Log-Rank~\cite{gehrmann2019gltr}, Fast-DetectGPT~\cite{bao2024fastdetectgpt}, Binoculars~\cite{hans2024binoculars}, DetectGPT~\cite{mitchell2023detectgpt}, and DNA-GPT~\cite{yang2024dnagpt}, on the DetectRL benchmark~\cite{wu2024detectrl}.
We compare \scheme{} with the detector of Zhou et al.~\cite{zhou2026beyondsemantic} on FaceForensics++ DFD face-swap videos (Table~\ref{tab:zhou-video-comparison}). The software baseline is trained on DFD; \scheme{} uses $480$ frames per video.
We compare \scheme{} with compression ratio, the size-only baseline, on identical samples.

\begin{table*}[!tb]
\centering

{\fontsize{8}{9.2}\selectfont\tablesmaller
\definecolor{tThreeLasso}{HTML}{F6E3E4}\definecolor{tThreeGrp}{gray}{0.93}%
\newcommand{\tThreeIn}{\arrayrulecolor{black!22}\hline\arrayrulecolor{black}}%
\setlength{\tabcolsep}{3pt}\renewcommand{\arraystretch}{1.12}
\begin{tabularx}{.99\textwidth}{>{\raggedright\arraybackslash}p{.18\textwidth}>{\centering\arraybackslash}p{.05\textwidth}>{\centering\arraybackslash}p{.085\textwidth}>{\centering\arraybackslash}p{.075\textwidth}>{\centering\arraybackslash}p{.085\textwidth}>{\centering\arraybackslash}p{.06\textwidth}>{\centering\arraybackslash}p{.08\textwidth}>{\centering\arraybackslash}p{.055\textwidth}>{\centering\arraybackslash}p{.07\textwidth}>{\centering\arraybackslash}X}
\Xhline{0.9pt}
\textbf{Detector} & \makecell{\textbf{Acc.}\\\textbf{(\%)}} & \makecell{\textbf{TPR/TNR}\\\textbf{(\%)}} & \makecell{\textbf{Latency}\\\textbf{(s)}} & \makecell{\textbf{Energy}\\\textbf{(J)}} & \makecell{\textbf{Power}\\\textbf{(W)}} & \textbf{RT/edge} & \makecell{\textbf{Gain}\\\textbf{(pp)}} & \textbf{Text det.} & \textbf{Signal}\\ \Xhline{0.6pt}
\rowcolor{tThreeGrp}\multicolumn{10}{l}{\textbf{FaceForensics++ DFD video} (per 480-frame video)}\\ \tThreeIn
Zhou et al.~(CPU)~\textcolor{blue}{\cite{zhou2026beyondsemantic}} & 49 & 67/29 & 8.77 & 1{,}315.5 & 150 & No/No & +18 & No & Software\\ \tThreeIn
Zhou et al.~(GPU)~\textcolor{blue}{\cite{zhou2026beyondsemantic}} & 49 & 67/29 & 0.627 & 158.864 & 252 & No/No & +18 & No & Software\\ \tThreeIn
\rowcolor{tThreeLasso}\textbf{\scheme{}} & \textbf{67} & \textbf{92/43} & \textbf{0.35} & \textbf{2.2} & \textbf{6.3} & \textbf{Yes/Yes} & --- & \textbf{Yes} & \textbf{Hardware}\\ \Xhline{0.6pt}
\rowcolor{tThreeGrp}\multicolumn{10}{l}{\textbf{DRCT-2M images} (per image)}\\ \tThreeIn
Zhou et al.~(CPU)~\textcolor{blue}{\cite{zhou2026beyondsemantic}} & 79 & 60/98 & 0.02 & 2.97 & 150 & Yes/No & --- & No & Software\\ \tThreeIn
Zhou et al.~(GPU)~\textcolor{blue}{\cite{zhou2026beyondsemantic}} & 79 & 60/98 & 0.0013 & 0.327 & 252 & Yes/No & --- & No & Software\\ \tThreeIn
CNNSpot~(CPU)~\textcolor{blue}{\cite{wang2020cnndetection}} & 50 & 0/99 & 0.016 & 2.44 & 150 & Yes/No & +19 & No & Software\\ \tThreeIn
CNNSpot~(GPU)~\textcolor{blue}{\cite{wang2020cnndetection}} & 50 & 0/99 & 0.0011 & 0.288 & 255 & Yes/No & +19 & No & Software\\ \tThreeIn
UnivFD~(CPU)~\textcolor{blue}{\cite{ojha2023univfd}} & 57 & 14/99 & 0.389 & 58.3 & 150 & No/No & +12 & No & Software\\ \tThreeIn
UnivFD~(GPU)~\textcolor{blue}{\cite{ojha2023univfd}} & 57 & 14/99 & 0.0102 & 2.545 & 251 & Yes/No & +12 & No & Software\\ \tThreeIn
\rowcolor{tThreeLasso}\textbf{\scheme{}} & \textbf{69} & \textbf{63/77} & \textbf{0.007} & $\boldsymbol{35{\times}10^{-3}}$ & \textbf{5.6} & \textbf{Yes/Yes} & --- & \textbf{Yes} & \textbf{Hardware}\\ \Xhline{0.9pt}
\end{tabularx}}
\vspace{2pt}\caption{Baselines on CPU (Xeon 5420+, RAPL) and GPU vs. \scheme{} on QAT; DFD cost per 480-frame video, DRCT-2M per image.}
\label{tab:zhou-lasso-comparison}\label{tab:zhou-video-comparison}
\end{table*}

\section{Results}
\label{sec:qatscan-detect}
\label{sec:threat-model}
\label{sec:real_vs_fake}
\subsection{How Accurate Is \scheme{}?}
\looseness=-1 \noindent\textbf{Detection results.} We report accuracy on each benchmark of Section~\ref{subsec:benchmarks}, with confusion matrices $[\mathrm{TP},\mathrm{FN};\mathrm{FP},\mathrm{TN}]$\,\% by true class.
On text, \scheme{} reaches $100\%$ ($[100,0;0,100]$) on WikiText-103 (3B-Instruct, natural format) (Threshold:- 31k jobs/s), FAIDSet (Gemini) (Threshold:- 31.8k jobs/s), RAID (GPT-2 XL)( Threshold:- 26.2k jobs/s), and DetectRL (back-translation, fixed cutoff) (27k jobs/s), against $100\%$ $75\%$, $79\%$, and $58\%$ for Binoculars~\cite{hans2024binoculars}; on Beemo (Llama P2, edited cutoff), it reaches $79\%$ ($[77,23;19,81]$) (Threshold:- 25.2k jobs/s) against $78\%$.

On images, it reaches $90\%$ on GenImage (BigGAN; $[85,15;5,95]$), $98\%$ on Self-Synthesis (BEGAN; $[98,2;2,98]$), and $95\%$ on DRCT-2M (SDXL-Refiner; $[95,5;5,95]$), against $98\%$, $97\%$, and $79\%$ for Zhou et al.~\cite{zhou2026beyondsemantic}; it also reaches $70\%$ on CLIC\slash DiffusionDB ($[59,41;19,81]$) (Threshold:- 14.1k jobs/s) and $77\%$ on SRD/GPT-4o receipts ($[92,8;39,61]$) (Threshold:- 14k jobs/s).
On code, it reaches $70\%$ on STACAD (Claude 3 Haiku; $[82,18;43,57]$).
On video, it reaches $67\%$ on FaceForensics++ DFD ($[92,8;57,43]$) and $78\%$ on FETV/MSR-VTT ($[78,22;21,79]$), against $49\%$ and $62\%$ for Zhou et al.

\looseness=-1 \noindent\textbf{Comparison with software text detectors.} We train \scheme{} on DetectRL's~\cite{wu2024detectrl} unmodified human and AI text, then classify unseen human text and AI text rewritten by DIPPER~\cite{krishna2023dipper} to resemble human writing; Table~\ref{tab:detectrl-baseline-cost} compares \scheme{} with five software detectors on detection quality, cost, and deployment.

\looseness=-1 \noindent\textbf{Unedited text.} \scheme{} achieves $1.000$ AUROC, ahead of Binoculars ($0.943$), Log-Rank ($0.883$), Fast-DetectGPT ($0.874$), DetectGPT ($0.786$), and DNA-GPT ($0.608$). Its advantage ranges from $5.7$ percentage points over Binoculars to $39.2$ over DNA-GPT, with $11.7$ over Log-Rank, $12.6$ over Fast-DetectGPT, and $21.4$ over DetectGPT.

\begin{figure}[!ht]
\centering\footnotesize\selectcolormodel{natural}
\definecolor{drctRed}{HTML}{C44E52}\definecolor{drctDark}{HTML}{7A2E33}\definecolor{drctGray}{HTML}{5A5A5A}
\begin{tikzpicture}[gainlbl/.style={font=\tiny,text=drctDark,inner sep=0pt}]
\begin{axis}[width=\columnwidth, height=3cm, scale only axis=false,
  xmin=-0.5,xmax=11.5, ymin=0,ymax=62, ytick={0,10,20,30,40,50},
  xtick={0,...,11}, xticklabels={{SDXL Refiner},{LCM-SDXL},{SDXL Turbo},{SDXL},{SD Turbo},{SDXL Ctrl},{SD v2},{SD v2 Ctrl},{SD v1 Ctrl},{SDXL DR},{SD v1 DR},{SD v2 DR}},
  xticklabel style={font=\scriptsize,rotate=40,anchor=north east,inner sep=1pt},
  tick label style={font=\scriptsize}, ylabel={Error rate (\%)}, label style={font=\footnotesize},
  ylabel style={yshift=-4pt},
  axis x line*=bottom, axis y line*=left, tick style={black}, major tick length=2pt,
  ymajorgrids=false, xmajorgrids=false, clip=false,
  legend style={font=\scriptsize,draw=none,fill=none,at={(0.5,1.02)},anchor=south,legend columns=2,
    /tikz/every even column/.append style={column sep=8pt}},
  legend cell align=left]
\draw[black!25,line width=0.3pt] (axis cs:-0.5,55) -- (axis cs:11.5,55);
\addplot[drctGray,thick,densely dashed,mark=diamond*,mark size=1.8pt,mark options={solid,fill=white}] coordinates {(0,6.5) (1,21.3) (2,12.3) (3,36.4) (4,26.8) (5,21.2) (6,36.0) (7,24.7) (8,28.5) (9,50.8) (10,50.7) (11,50.8)};
\addlegendentry{Zhou et al.~\cite{zhou2026beyondsemantic} error}
\addplot[drctRed,thick,mark=*,mark size=1.4pt,mark options={fill=drctRed,draw=drctDark},
  error bars/.cd,y dir=both,y explicit,error bar style={drctDark,line width=0.5pt},error mark options={rotate=90,mark size=1.3pt,line width=0.5pt,drctDark}] coordinates {
 (0,4.8) +=(0,0) -=(0,0.1)
 (1,9.7) +=(0,0.3) -=(0,0.2)
 (2,11.7) +=(0,0.6) -=(0,0.6)
 (3,13.2) +=(0,0) -=(0,0.1)
 (4,17.2) +=(0,0.7) -=(0,0.6)
 (5,19.6) +=(0,1.5) -=(0,1.6)
 (6,21.5) +=(0,1.0) -=(0,1.1)
 (7,23.5) +=(0,0.4) -=(0,0.5)
 (8,25.1) +=(0,0.4) -=(0,0.4)
 (9,29.8) +=(0,1.0) -=(0,0.9)
 (10,32.0) +=(0,3.7) -=(0,3.7)
 (11,33.1) +=(0,1.8) -=(0,1.8)};
\addlegendentry{\scheme{} error (whiskers: FPR, FNR)}
\node[gainlbl] at (axis cs:0,62.5) {1.7};
\node[gainlbl] at (axis cs:1,62.5) {11.6};
\node[gainlbl] at (axis cs:2,62.5) {0.6};
\node[gainlbl] at (axis cs:3,62.5) {23.2};
\node[gainlbl] at (axis cs:4,62.5) {9.6};
\node[gainlbl] at (axis cs:5,62.5) {1.6};
\node[gainlbl] at (axis cs:6,62.5) {14.5};
\node[gainlbl] at (axis cs:7,62.5) {1.2};
\node[gainlbl] at (axis cs:8,62.5) {3.4};
\node[gainlbl] at (axis cs:9,62.5) {21.0};
\node[gainlbl] at (axis cs:10,62.5) {18.7};
\node[gainlbl] at (axis cs:11,62.5) {17.7};
\node[gainlbl,anchor=east] at (axis cs:-0.6,62.5) {Gain (pp)};
\end{axis}
\end{tikzpicture}
    \caption{DRCT-2M error for 12 generators, sorted by \scheme{} error; top row: accuracy gain over Zhou et al.~\cite{zhou2026beyondsemantic}.}
    \label{fig:image-video-results}\label{tab:res-image}\label{tab:res-image-drct}
\end{figure}
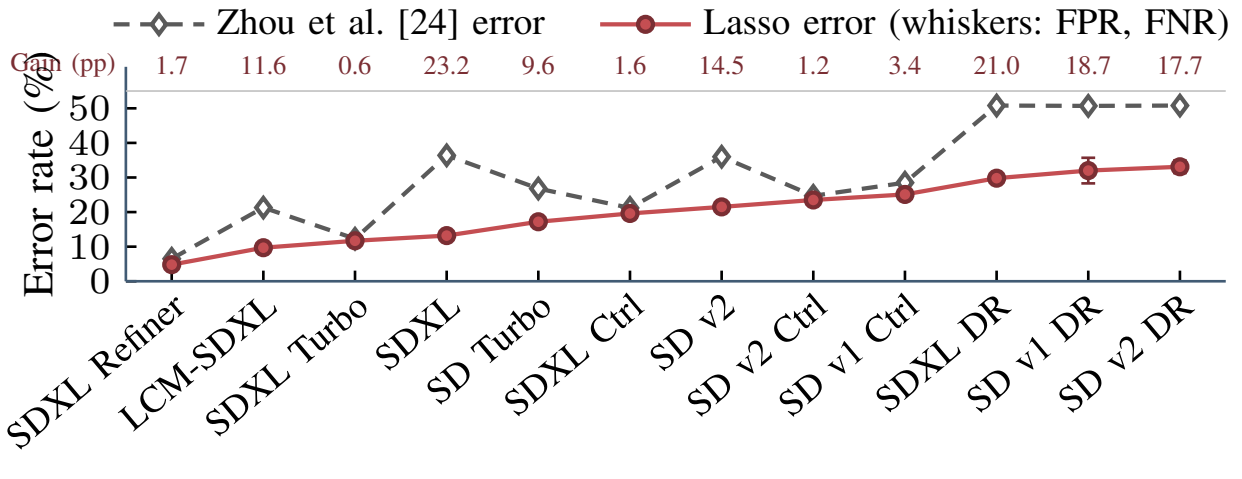
\looseness=-1 \noindent\textbf{Image detection.} Figure~\ref{fig:image-video-results} reports DRCT-2M results for $12$ generators, sorted by \scheme{} error, with the gain over Zhou et al. above each generator; detection strength varies by generator.

\subsection{Does Detection Survive Evasion?}\label{subsec:code-detection}
\looseness=-1 \noindent\textbf{Robustness to paraphrasing.} %
Under DIPPER paraphrasing, \scheme{} retains a mean AUROC of $0.971$, a reduction of only $2.9$ points. Binoculars reaches $0.808$, followed by DNA-GPT, Fast-DetectGPT, and Log-Rank at $0.677$--$0.742$; DetectGPT falls to $0.462$.
\scheme{} leads every baseline by at least $16.3$ percentage points, with margins of $23.9$ over Fast-DetectGPT, $29.4$ over Log-Rank, and $50.9$ over DetectGPT.

\looseness=-1 \noindent\textbf{Robustness to perturbation.} Under character-, word-, and sentence-level perturbation, \scheme{} again leads, with a mean AUROC of $0.849$. Binoculars and DNA-GPT reach $0.769$ and $0.674$; Fast-DetectGPT, Log-Rank, and DetectGPT fall below $0.5$, worse than chance.
\scheme{} leads Binoculars by $8.0$ points and DNA-GPT by $17.5$, and the margins widen to $39.0$ over Fast-DetectGPT, $45.5$ over Log-Rank, and $71.7$ over DetectGPT.

\looseness=-1 \noindent\textbf{Robustness across rewriting attacks.}\label{subsec:transformation-evasion} Figure~\ref{fig:roc-sweep} extends the evaluation to six rewriting attacks: back-translation, DIPPER paraphrasing, polishing, and character, word, and sentence perturbation. The threshold is fitted on unedited human and AI text ($27{,}033.9$~jobs/s), held fixed, and applied to a disjoint test split; AUROC measures class ordering independently of this threshold.

\begin{figure*}[!htbp]
\centering
\resizebox{\textwidth}{!}{%
\selectcolormodel{natural}\definecolor{lassoRed}{HTML}{C44E52}%
\begin{tikzpicture}
\pgfplotsset{raidcompact/.style={scale only axis,height=1.5cm,ybar=0.4pt,bar width=2.8pt,ymin=0,ymax=105,ytick={0,50,100},axis line style={black,line width=.5pt},tick style={black,line width=.4pt},ymajorgrids=false,grid style={densely dashed,black!45,line width=.3pt},tick label style={font=\fontsize{7}{8}\selectfont},xticklabel style={rotate=45,anchor=east},title style={font=\fontsize{8}{9}\selectfont,yshift=1pt},enlarge x limits=false,clip=false}}
\begin{axis}[raidcompact,at={(0cm,0)},anchor=south west,width=10.15cm,xlabel={(a) Generator transfer},xlabel style={font=\fontsize{8}{9}\selectfont,yshift=-2pt},xmin=-.6,xmax=11.6,xtick={0,1,2,3,4,5,6,7,8,9,10,11},xticklabels={MPT chat,LLaMA-2,GPT-2 XL,GPT-3,Cohere chat,Mistral chat,MPT-30B,ChatGPT,GPT-4,Cohere,Mistral-7B,Mean},ylabel={Accuracy (\%)},ylabel style={font=\fontsize{8}{9}\selectfont},legend style={at={(.86,1.07)},anchor=south,legend columns=5,font=\fontsize{8}{9}\selectfont,draw=black,fill=white,column sep=7pt},]
\addplot[area legend,fill=lassoRed,draw=black,line width=.25pt] coordinates {(0,93.8) (1,93.8) (2,100) (3,87.5) (4,87.5) (5,87.5) (6,87.5) (7,87.5) (8,93.8) (9,62.5) (10,75) (11,86.9)};
\addlegendentry{\scheme{}: seen generator}
\addplot[area legend,fill=lassoRed!40,draw=black,line width=.25pt] coordinates {(0,93.8) (1,93.8) (2,87.5) (3,87.5) (4,87.5) (5,87.5) (6,87.5) (7,81.2) (8,75) (9,75) (10,50) (11,82.4)};
\addlegendentry{\scheme{}: unseen generator}
\addplot[area legend,fill=white,draw=black,line width=.25pt] coordinates {(0,75) (1,56.2) (2,75) (3,81.2) (4,68.8) (5,68.8) (6,75) (7,75) (8,50) (9,37.5) (10,68.8) (11,66.5)};
\addlegendentry{IAA decomp.: seen generator}
\addplot[area legend,pattern=north east lines,draw=black,line width=.25pt] coordinates {(0,62.5) (1,62.5) (2,75) (3,81.2) (4,68.8) (5,81.2) (6,75) (7,75) (8,43.8) (9,37.5) (10,56.2) (11,65.3)};
\addlegendentry{IAA decomp.: unseen generator}
\addplot[area legend,fill=black!18,draw=black,line width=.25pt] coordinates {(0,97.2) (1,97.7) (2,79.9) (3,96.7) (4,90.8) (5,97.3) (6,72.7) (7,96.7) (8,75.1) (9,86.5) (10,70.4) (11,87.4)};
\addlegendentry{Binoculars}
\end{axis}
\begin{axis}[raidcompact,at={(10.6cm,0)},anchor=south west,width=6.7cm,xlabel={(b) Domain transfer},xlabel style={font=\fontsize{8}{9}\selectfont,yshift=-2pt},xmin=-.6,xmax=7.6,xtick={0,1,2,3,4,5,6,7},xticklabels={Wikipedia,Books,Poetry,Reviews,Reddit,News,Abstracts,Recipes},yticklabels={},]
\addplot[area legend,fill=lassoRed,draw=black,line width=.25pt] coordinates {(0,89.3) (1,82.4) (2,84.6) (3,81.2) (4,75) (5,75) (6,83.3) (7,97.1)};
\addplot[area legend,fill=lassoRed!40,draw=black,line width=.25pt] coordinates {(0,89.3) (1,85.3) (2,84.6) (3,81.2) (4,80) (5,77.5) (6,50) (7,50)};
\addplot[area legend,fill=white,draw=black,line width=.25pt] coordinates {(0,64.3) (1,61.8) (2,80.8) (3,62.5) (4,45) (5,75) (6,66.7) (7,55.9)};
\addplot[area legend,pattern=north east lines,draw=black,line width=.25pt] coordinates {(0,67.9) (1,61.8) (2,69.2) (3,68.8) (4,45) (5,60) (6,50) (7,52.9)};
\addplot[area legend,fill=black!18,draw=black,line width=.25pt] coordinates {(0,86.4) (1,92.5) (2,89) (3,88.8) (4,87.1) (5,86.1) (6,85.8) (7,86.9)};
\end{axis}
\end{tikzpicture}%
}
\caption{RAID transfer to unseen generators and domains, no pooling; one frozen threshold (26,401.2 jobs/s) on QAT and (29,460 jobs/s) IAA}
\label{fig:raid-transfer}\label{tab:raid-stacad-results}
\end{figure*}
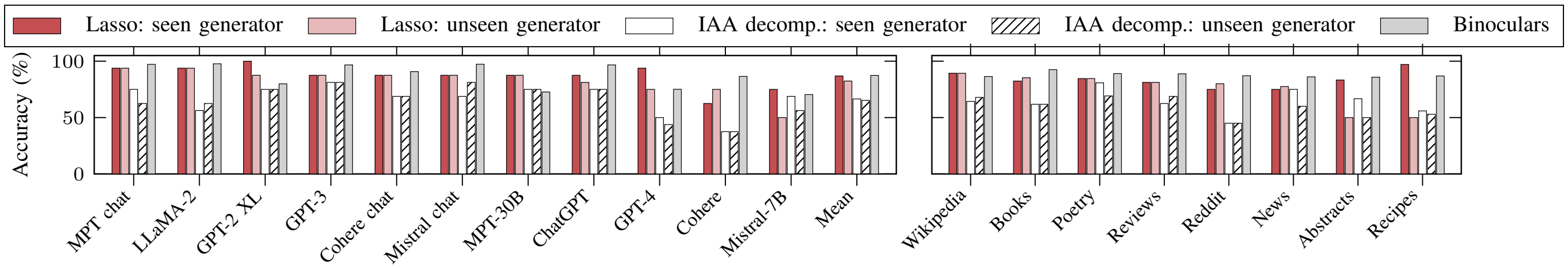

\subsection{Does Detection Generalize to Unseen Data?}

\looseness=-1 \noindent\textbf{Detection across generators and domains.} We evaluate \scheme{} on RAID~\cite{raid}, covering $11$ generators and eight text domains (Figure~\ref{fig:raid-transfer}).

\looseness=-1 \noindent\textbf{Generator transfer.} Trained on one generator(Llama-2) and tested on another, without pooling examples,~\scheme{} reaches $82\%$ mean accuracy on QAT (Figure~\ref{fig:raid-transfer}a), detecting generators absent from training (Threshold:- 26.4k jobs/s). On IAA, \scheme{} reached 66.5\% mean accuracy on similar configuration (Threshold:- 29.4k jobs/s).

\looseness=-1 \noindent\textbf{Domain transfer.} With the test domain excluded from calibration, mean accuracy across the eight domains is $74\%$ using QAT and $64\%$ using IAA, so the signal transfers across domains (Figure~\ref{fig:raid-transfer}b).

\looseness=-1 \noindent\textbf{Transfer to unseen generators.} On WikiText, excluding the tested generator from calibration retains $95\%$ accuracy for instruction-tuned models with matched formatting (Threshold:- 28.8k jobs/sec) and $100\%$ with natural-text formatting (Threshold:- 31.2k jobs/s); the separation extends to unseen generators.

\subsection{How Much Input Does Detection Need?}

\looseness=-1 \noindent\textbf{Text observation size study.} For WikiText/GPT-2, accuracy rises from $67\%$ at $512$\,B to $85\%$ at $64$\,KB and $98\%$ at $256$\,KB ($0.996$ AUROC; Figure~\ref{fig:prelim-silicon}).

\looseness=-1 \noindent\textbf{More frames per video.} Keeping clips actor-disjoint and the sampling rate fixed, we increase the number of observed frames. %
Logistic-regression accuracy is $62.9\%$ with $30$ frames, dips to $59.6\%$ at $60$, rises to $64.1\%$ at $160$, and then plateaus near $65\%$ ($65.4\%$ at $480$); gradient boosting follows the same trend ($60.5\%$ to $64.1\%$ by $160$ frames, then $63.6$--$66.1\%$). Thus, $160$ frames suffice.

\subsection{Is \scheme{} Real-Time and Energy-Efficient?}\label{subsec:images-video}
\looseness=-1 \noindent\textbf{Detection latency.} \scheme{} takes $197$\,$\mu$s per check. Log-Rank and Fast-DetectGPT take $70$ and $71$~ms, Binoculars takes $234$~ms, and DNA-GPT and DetectGPT take $3.75$ and $17.4$~s (Table~\ref{tab:detectrl-baseline-cost}).
\scheme{} is therefore $356\times$ faster than Log-Rank, $359\times$ faster than Fast-DetectGPT, and $1{,}188\times$ faster than Binoculars, reaching $19{,}014\times$ over DNA-GPT and $88{,}231\times$ over DetectGPT.

\looseness=-1 \noindent\textbf{Real-time latency.} For a real-time target of $100$~ms per document, \scheme{}, Log-Rank, and Fast-DetectGPT meet the target, while Binoculars, DNA-GPT, and DetectGPT exceed it; \scheme{}'s $197\,\mu$s is more than $500\times$ below the limit.

\looseness=-1 \noindent\textbf{Energy and edge power.} \scheme{} draws the
$6.3$-W QAT power of Table~\ref{tab:zhou-video-comparison} for its
$0.35$ s check, or $1.2\times10^{-3}$~J, compared with
$16.3$--$24.9$~J for Fast-DetectGPT, Log-Rank, and Binoculars and $369$ and
$3{,}258$~J for DNA-GPT and DetectGPT (Table~\ref{tab:detectrl-baseline-cost}). Thus \scheme{} uses over $99\%$ less
energy per check than every baseline, and at ten checks per second it
averages $12$~mW, well within a $5$--$6$-W edge budget, whereas the lowest-energy
baseline ($16.3$~J) needs $163$~W. This low power and energy consumption, makes \scheme{} highly sustainable to increasing demand of data provenance.

\looseness=-1 \noindent\textbf{Detector complexity.} \scheme{} classifies accelerator measurements without language-model inference, whereas all five software baselines depend on language-model computation, which increases the work per check.

\looseness=-1 \noindent\textbf{Video detection latency.} With three repetitions, each frame requires $18$ QAT requests at $40.5\,\mu$s each, $0.729$\,ms in total, giving $1{,}372$ frames/s of compression capacity.

\looseness=-1 \noindent{\setlength{\fboxsep}{0pt}\colorbox{revhl}{\textbf{Quantitative comparison.}}} On DFD video, \scheme{} reaches $0.739$ AUROC and $67\%$ accuracy, compared with $0.474$ AUROC and $49\%$ accuracy for Zhou et al., gains of $26.5$ percentage points in AUROC and $18$ in accuracy. \scheme{} also detects generated text, extending its use beyond the image-based detector. Per video, \scheme{} uses $72\times$ less energy and has $1.8\times$ lower latency than Zhou et al. on the A100 GPU ($597\times$ and $25\times$ against the CPU); its $0.35$-s latency is $480$ frames at $0.729$~ms each. On DRCT-2M images, it uses $9.3\times$ less energy than the GPU and $85\times$ less than the CPU; its per-image latency is $2.9\times$ lower than the CPU's but $5.4\times$ higher than the GPU's (Table~\ref{tab:zhou-video-comparison}). These measurements compare a full A100 GPU with a single QAT instance.

\subsection{Does Timing Beat Compression Ratio?}
\label{subsec:fixed-ratio-dipper}
\label{sec:ratio-comparison}

\looseness=-1 \noindent\textbf{Matched-resolution image encoding.} With resolution matched and the score direction fixed after calibration, \scheme{} improves AUROC over compression ratio by $6.6$--$32.1$ percentage points ($11$--$57\%$ relative) across the seven tested image generators; for SD~v1.4, AUROC rises from $0.659$ to $0.980$, a $49\%$ relative gain.

\looseness=-1 \noindent\textbf{Video detection with matched codecs.} Across videos matched by codec, frame count, resolution, and frame rate, with score direction calibrated on another generator, \scheme{} improves AUROC over compression ratio by $13.2$ points for MPEG-4 ($0.597$ to $0.729$, $22\%$ relative) and $8.9$ for H.264 ($0.581$ to $0.670$, $15\%$).

\looseness=-1 \noindent\textbf{Detection at fixed compression ratio.} \scheme{} retains $66\%$ accuracy on DIPPER-rewritten~\cite{krishna2023dipper} text when compression ratio is held fixed at it's native 1.28KB per document size, showing that timing carries information beyond output size.
The token-cost model explains why: compressed size sums the tokens' code lengths, but decode work depends on the literal count, match count, and bytes copied (Equation~\ref{eq:qre-costmodel}). A fixed compression ratio leaves the literal/match split and the match lengths free, so blocks with equal ratios can differ in decode work, which timing resolves and size does not.

\begin{figure*}[!t]
\centering\selectcolormodel{natural}
\definecolor{noisePurple}{HTML}{7A2E33}
\definecolor{noiseRed}{HTML}{C44E52}
\begin{tikzpicture}
\begin{groupplot}[
  group style={group size=4 by 1, horizontal sep=14pt},
  scale only axis, width=0.195\textwidth, height=1.5cm,
  xmin=0,xmax=14,xtick={0,4,8,14},
  tick label style={font=\footnotesize}, label style={font=\footnotesize},
  title style={font=\small\bfseries, yshift=-5pt},
  tick style={black},axis line style={black},
  ymajorgrids=false,
  ymin=.4,ymax=1.1,ytick={.5,.7,.9},
  enlarge x limits=0.03,
  every axis plot/.append style={thick,mark=*,mark size=1.1pt},
  legend style={font=\scriptsize,draw=none,fill=none,inner sep=0.5pt,row sep=-2pt,
                /tikz/every even column/.append style={column sep=2pt}},
  legend cell align=left,
  legend image code/.code={\draw[mark repeat=2,mark phase=2,#1] plot coordinates {(0cm,0cm) (0.15cm,0cm) (0.3cm,0cm)};},
]
\nextgroupplot[title={(a) Text}, ylabel={AUROC}, ylabel style={yshift=-3pt}]
\addplot[gray!55,densely dotted,thin,mark=none,forget plot] coordinates {(0,.5)(14,.5)};
\addplot[noiseRed,mark options={fill=noiseRed},error bars/.cd,y dir=both,y explicit,error bar style={noiseRed!80!black,line width=0.5pt},error mark options={rotate=90,mark size=1.2pt,line width=0.5pt}] coordinates {
 (0,0.914) +=(0,0.061) -=(0,0.078)
 (1,0.688) +=(0,0.122) -=(0,0.123)
 (2,0.713) +=(0,0.107) -=(0,0.106)
 (4,0.825) +=(0,0.161) -=(0,0.107)
 (6,0.715) +=(0,0.187) -=(0,0.164)
 (8,0.712) +=(0,0.136) -=(0,0.139)
 (10,0.672) +=(0,0.102) -=(0,0.171)
 (12,0.542) +=(0,0.167) -=(0,0.085)
 (14,0.469) +=(0,0.104) -=(0,0.083)};
\nextgroupplot[title={(b) Images}, yticklabels={}]
\addplot[gray!55,densely dotted,thin,mark=none,forget plot] coordinates {(0,.5)(14,.5)};
\addplot[noisePurple,densely dashed,mark=square*,mark options={solid,fill=white}] coordinates {(0,0.968) (1,0.967) (2,0.967) (4,0.967) (6,0.967) (8,0.967) (10,0.966) (12,0.956) (14,0.956)};
\addplot[black] coordinates {(0,0.868) (1,0.866) (2,0.867) (4,0.867) (6,0.866) (8,0.867) (10,0.861) (12,0.854) (14,0.842)};
\addplot[noiseRed,densely dotted,mark=triangle*,mark options={solid}] coordinates {(0,0.841) (1,0.838) (2,0.842) (4,0.84) (6,0.839) (8,0.841) (10,0.831) (12,0.821) (14,0.82)};
\nextgroupplot[title={(c) Video}, yticklabels={}]
\addplot[gray!55,densely dotted,thin,mark=none,forget plot] coordinates {(0,.5)(14,.5)};
\addplot[gray,densely dashed,thin,mark=none,forget plot] coordinates {(0,.805)(14,.805)};
\addplot[noisePurple,densely dashed,mark=square*,mark options={solid,fill=white}] coordinates {(0,0.785) (1,0.824) (2,0.769) (4,0.82) (6,0.802) (8,0.835) (10,0.828) (12,0.79) (14,0.82)};
\addplot[black] coordinates {(0,0.745) (1,0.796) (2,0.73) (4,0.783) (6,0.766) (8,0.804) (10,0.799) (12,0.771) (14,0.793)};
\addplot[noiseRed,densely dotted,mark=triangle*,mark options={solid}] coordinates {(0,0.705) (1,0.768) (2,0.692) (4,0.746) (6,0.731) (8,0.774) (10,0.77) (12,0.753) (14,0.765)};
\nextgroupplot[title={(d) QAT load}, ymin=-5,ymax=108,ytick={0,50,100},
  yticklabel pos=right, ylabel={Util.\ (\%)}, ylabel style={yshift=3pt},
  legend columns=1, legend style={at={(0.97,0.45)},anchor=east}]
\addplot[noiseRed,densely dotted,mark=triangle*,mark options={solid}] coordinates {(0,0.0) (1,56.0) (2,69.0) (4,100.0) (6,100.0) (8,100.0) (10,100.0) (12,100.0) (14,100.0)};
\addlegendentry{Shared}
\addplot[black] coordinates {(0,0.0) (1,0.0) (2,0.0) (4,0.0) (6,0.0) (8,0.0) (10,0.0) (12,0.0) (14,0.0)};
\addlegendentry{Idle}
\end{groupplot}
\path (group c2r1.south east) -- (group c3r1.south west) coordinate[midway] (noiseXAxisCenter);
\node[font=\footnotesize, anchor=north] at ([yshift=-11pt]noiseXAxisCenter) {Co-tenant compression threads};
\node[font=\scriptsize, anchor=north, inner sep=0pt] at ([yshift=-22pt]noiseXAxisCenter) {%
 (b,\,c):\enspace
 \tikz[baseline=-0.6ex]{\draw[noisePurple,thick,densely dashed] (0,0)--(0.42,0); \node[anchor=center,solid,draw=noisePurple,thick,fill=white,inner sep=0pt,minimum size=3pt,rectangle] at (0.21,0){};}\,Refiner\,/\,T2V\enspace
 \tikz[baseline=-0.6ex]{\draw[black,thick] (0,0)--(0.42,0); \node[anchor=center,fill=black,circle,inner sep=0pt,minimum size=3pt] at (0.21,0){};}\,Mean\enspace
 \tikz[baseline=-0.6ex]{\draw[noiseRed,thick,densely dotted] (0,0)--(0.42,0); \fill[noiseRed] (0.21,0) ++(90:2.1pt) -- ++(-120:3.6pt) -- ++(0:3.6pt) -- cycle;}\,LCM\,/\,ModelScope};
\end{tikzpicture}
\caption{AUROC under co-tenant QAT load; threshold frozen at $N{=}0$.}
\label{fig:noise}
\end{figure*}
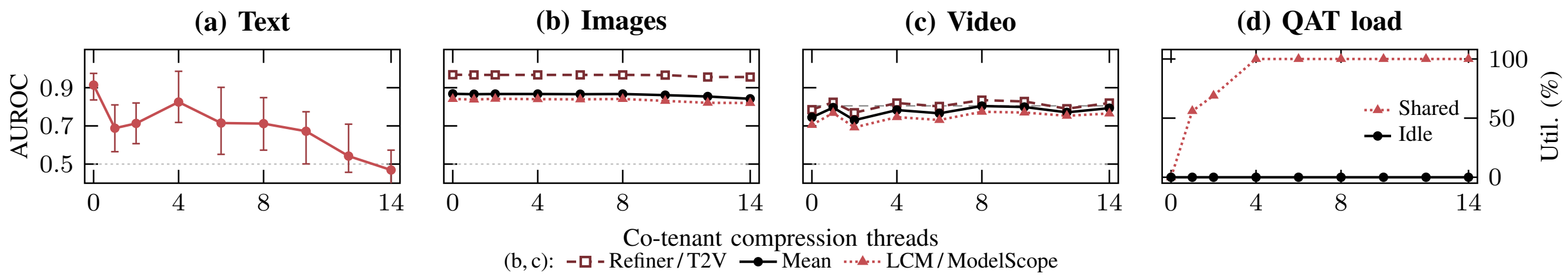

\subsection{Does Detection Survive Co-tenant Load?}
\label{subsec:noise-analysis}

\looseness=-1 On text (DetectRL), the mean AUROC is $0.914$ on the idle device and remains at or above $0.67$ through $N{=}10$ (Figure~\ref{fig:noise}a). On images (DRCT-2M), the mean AUROC moves only from $0.868$ on the idle device to $0.842$ at full saturation (Figure~\ref{fig:noise}b). On video (FETV/MSR-VTT), the mean AUROC of the video detector, which combines many frames, ranges from $0.73$ to $0.80$ (T2V $0.77$--$0.84$, ModelScope $0.69$--$0.77$), with no trend in $N$ (Figure~\ref{fig:noise}c).

Figure~\ref{fig:noise}d reads the compression-slice utilization counter (\texttt{util\_cpr0}) during the sweep: on the shared device it rises $0\!\rightarrow\!56\!\rightarrow\!69\!\rightarrow\!100\%$ and remains pinned at $100\%$ for $N\geq4$. Saturation is thus established from the accelerator's own telemetry, not inferred from timing.

\subsection{Does Detection survive Adaptive attack ?}

\begin{figure}[t]
\centering
\includegraphics[width=0.9\linewidth]{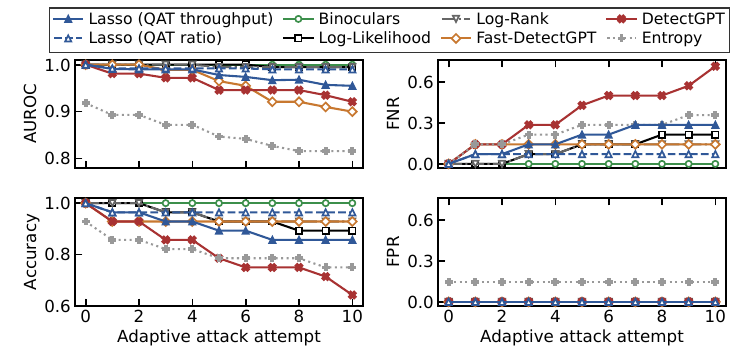}
\caption{Detector-aware adaptive attack on RAID (64\,KB bundles, 14 human /
14 AI; thresholds frozen; BERTScore $\geq$ 0.945). Top: AUROC and
false-negative rate; bottom: detection accuracy and false-positive rate.}
\label{fig:adaptive}
\end{figure}

We also attack each detector with its own detector-aware loop on the RAID
dataset, using 64\,KB bundles (14 human and 14 AI;
Figure~\ref{fig:adaptive}). For each AI bundle, the attacker splits the text into 3-sentence windows and
ranks them by how AI-like each looks to the target detector.
For Lasso, the ranking uses the fraction of each window's bytes that the
compressor encodes as copies of earlier text: windows with the fewest copies
produce the most literals, which slow QAT, so they are rewritten first. The attacker paraphrases the most AI-like 40\% of windows with DIPPER,
discards any rewrite whose BERTScore to the original falls below 0.945,
queries the real detector, and keeps the most human-like candidate; after a
round with no improvement, it widens the rewritten fraction by 20\%, for 10
attempts in total. Thresholds stay frozen and human text is never rewritten, so the
false-positive rate does not change.
After 10 attempts, Lasso's AUROC falls from 1.00 to 0.96 and 4 of the 14 AI
bundles cross its threshold.
DetectGPT (0.92; 10 of 14), Fast-DetectGPT (0.90; 2 of 14) and Entropy
(0.82; 5 of 14) lose more AUROC. To evade Lasso, a rewrite must make the text repeat itself more within the
reach of QAT's match finder, so that fewer of its bytes become literals
(Section~\ref{sec:qre-evidence-6}); under the meaning constraint, the
attacker achieves this for some bundles but not for most.

\section{Related Work}
\label{sec:related}

\vspace{0.5em}
\subsection{Text and Software Code Detection}\label{subsec:related-text-code}DetectGPT~\cite{mitchell2023detectgpt} scores a passage against about 100 perturbations generated by an LLM, which requires about 200 LLM inferences; in our DetectRL measurements it consumes $3{,}258$~J and $17.4$~s per document (Table~\ref{tab:detectrl-baseline-cost}). Fast-DetectGPT~\cite{bao2024fastdetectgpt} replaces the perturbations with conditional token sampling, reducing the cost to $16.3$~J and $71$~ms.
Binoculars~\cite{hans2024binoculars} compares how predictable a passage is under two LLMs ($24.9$~J, $234$~ms), and DNA-GPT~\cite{yang2024dnagpt} regenerates the second half of a truncated passage and compares it with the original ($369$~J, $3.75$~s). Each detector runs LLM inference for every check, and each depends on a scoring model close to the source model or on calibration to the target dataset. \scheme{} instead times how QAT compresses the document: it scores 64\,KB blocks, and it requires $1.2\times10^{-3}$~J and $197$~\textmu s per check, over $99\%$ less energy than every software baseline.

Paraphrasing with DIPPER~\cite{krishna2023dipper} weakens DetectGPT, Fast-DetectGPT, and Binoculars, reducing DetectGPT's rate at $1\%$ FPR from $70.3\%$ to $4.6\%$~\cite{krishna2023dipper}. \scheme{} retains $0.971$ AUROC on DIPPER-paraphrased 64\,KB blocks (Table~\ref{tab:detectrl-baseline-cost}).

\subsection{Image and Video Detection}\label{subsec:related-image-video}DIRE~\cite{ref:dire} reconstructs an image with a pretrained diffusion model and detects generation from the reconstruction difference. Zhou et al.~\cite{zhou2026beyondsemantic} pre-process pixels to emphasize generation artifacts before neural classification, which improves generalization on fully generated images. Applied frame by frame to FaceForensics++~\cite{roessler2019faceforensicspp} DFD face-swap videos, it reaches $49\%$ accuracy, against $67\%$ for \scheme{} (Table~\ref{tab:zhou-video-comparison}). These approaches inspect image content; \scheme{} measures how the input changes accelerator execution.
Zhou et al. report running on eight NVIDIA RTX 3090 GPUs, whereas \scheme{} uses the QAT accelerator integrated in the Xeon. In our measurements (Table~\ref{tab:zhou-video-comparison}), \scheme{} uses $72\times$ less energy and has $1.8\times$ lower latency per DFD video than Zhou et al. on an A100 GPU; on DRCT-2M images it uses $9.3\times$ less energy than the GPU but has $5.4\times$ higher latency. Zhou et al. is designed for images and does not address text or code, whereas \scheme{} detects AI-generated text and code.

Deepfake CLI~\cite{bhilare2022deepfakecli} accelerates two CNN detectors
that are too slow for live use, Eff-YNet~\cite{tjon2021effynet}
(19M-parameter encoder) and U-Net~\cite{ronneberger2015unet}
(31M parameters), by running them in int8 on a Xeon host with an AMD
VCK5000 Versal FPGA card, cutting latency by $54.5\%$ relative to a
Tesla~T4 GPU. \scheme{} processes $1{,}372$ frames per second on the one
QAT device of a Xeon Gold 5420+, $4.3\times$ CLI's $316.8$; 4th-generation
Xeon packages integrate up to four QAT endpoints, and assuming linear
scaling across them, the projected rate is $5{,}487$ frames per second
($17\times$) with no added hardware. Even when charged the full $132$-W CPU
package power rather than the $6.3$-W QAT power
(Table~\ref{tab:zhou-video-comparison}), \scheme{} draws $41.3\%$ less power
than CLI's $225$-W card scenario~\cite{amd2023vck5000power} and uses
$69.5$--$86.5\%$ less energy per frame ($0.096$ vs.\ $0.316$--$0.710$~J).

CLI evaluates one artifact type on one dataset; \scheme{} detects
AI-generated images, code, text, and video across $13$ datasets
(Section~\ref{subsec:benchmarks}. 
More fundamentally,
CLI speeds up an existing detector with established techniques
(quantization and FPGA offload), whereas \scheme{} establishes a hardware
observation channel for AI provenance, that is, whether an artifact was
produced by an AI model or by a human. This channel is a starting point for
finding processor components that respond to AI-generated artifacts, and
for processors that detect provenance through their own execution, for
example a CPU that acts as a root of trust against AI-generated fakes.

\section{Conclusion}
\label{sec:conclusion}

Within fifteen years, every person should be able to hold a
cheap, fast device, whether a phone, a card, or a pair of glasses, that tells them in
real time whether the text they read or the image or video they watch is real or machine-made. 

\scheme{} connects QAT's compression pipeline to data provenance. QAT's match
finder decides which repeats become back-references through a limited-capacity
candidate table, a 48-byte comparator, and an extender; its mandatory
verification (decompression) pass then charges a fixed cost per literal and per
match. Human-made and AI-generated artifacts repeat differently, so they produce
different token streams; therefore, QAT throughput reveals their origin. Intel
IAA shows the same effect at a smaller scale: its match finder reaches only
4\,KiB back, so it supports a weaker detector. This connection provides a
starting point for a careful study that identifies all the microarchitectural
components that respond to AI artifacts. Once these components are fully
understood, they can be used to build a custom detector inside a CPU core,
enabling the CPU to flag the origin of the data it processes on the fly. We
envision provenance checking as a by-product of ordinary computation: as a
program reads, compresses, or moves a file, the hardware flags AI-generated
content without separate tools.

\smallskip\noindent\textbf{AI Acknowledgments.} We used Perplexity Computer to assist with drafting and restructuring the manuscript. The authors take full responsibility for its content.

\bibliographystyle{IEEEtran}
\bibliography{qat_ref}

 \end{document}